\documentclass[showpacs,reprint,a4paper]{revtex4-1}

\usepackage{amsmath}
\usepackage{amssymb}
\usepackage{graphicx}
\usepackage{tikz}
\usepackage{slashed}
\usepackage{multirow}
\usepackage{hhline}
\usepackage{subcaption}
\usepackage{caption}
\usepackage{float}
\usetikzlibrary{external}
\immediate\write18{mkdir -p pgf-img}
\begin{document}

\title{Universal renormalization group flow toward perfect Fermi-surface nesting driven by enhanced electron-electron correlations in monolayer vanadium diselenide}
\author{Iksu Jang$^{1}$, Ganbat Duvjir$^{2}$, Byoung Ki Choi$^{3}$, Jungdae Kim$^{2}$, Young Jun Chang$^{3}$, and Ki-Seok Kim$^{1}$}
\affiliation{$^{1}$Department of Physics, POSTECH, Pohang, Gyeongbuk 790-784, Korea \\ $^{2}$Department of Physics, BRL, and EHSRC, University of Ulsan, Ulsan 44610, Korea \\ $^{3}$Department of Physics, University of Seoul, Seoul 02504, Korea}
\date{\today}

\begin{abstract}
Reducing thickness of three dimensional samples on appropriate substrates is a promising way to control electron-electron interactions, responsible for so called electronic reconstruction phenomena. Although the electronic reconstruction has been investigated both extensively and intensively in oxide heterostructure interfaces, this paradigm is not well established in the van der Waals hetero-interface system, regarded to be important for device applications. In the present study we examine nature of a charge ordering transition in monolayer vanadium diselenide ($VSe_{2}$), which would be distinguished from that of $VSe_{2}$ bulk samples, driven by more enhanced electron-electron correlations. We recall that $VSe_{2}$ bulk samples show a charge density wave (CDW) transition around $T_{CDW} \sim 105$ $K$, expected to result from Fermi surface nesting properties, where the low temperature CDW state coexists with itinerant electrons of residual Fermi surfaces. Recently, angle resolved photoemission spectroscopy measurements [Nano Lett. \textbf{18}, 5432 (2018)] uncovered that the Fermi surface nesting becomes perfect, where the dynamics of hot electrons is dispersionless along the orthogonal direction of the nesting wave-vector. In addition, scanning tunneling microscopy measurements [Nano Lett. \textbf{18}, 5432 (2018)] confirmed that the resulting CDW state shows essentially the same modulation pattern as the three dimensional system of $VSe_{2}$. Here, we perform the renormalization group analysis based on an effective field theory in terms of critical CDW fluctuations and hot electrons of imperfect Fermi-surface nesting. As a result, we reveal that the imperfect nesting universally flows into perfect nesting in two dimensions, where the Fermi velocity along the orthogonal direction of the nesting vector vanishes generically. We argue that this electronic reconstruction is responsible for the observation [Nano Lett. \textbf{18}, 5432 (2018)] that the CDW transition temperature is much more enhanced to be around $T_{CDW} \sim 350$ $K$ than that of the bulk sample.
\end{abstract}

\maketitle

\section{Introduction}

Strongly correlated electrons had been expected to realize in transition metal dichalcogenides (TMDCs), involved with the quasi two dimensional lattice structure \cite{TMDC_Review}. Actually, several compounds of the TMDC family have shown physics of strong correlations, for example, local-moment signatures in Ir-dichalcogenides \cite{LM_Ir_TMDC} and Mott insulating physics in transition metal sulfides \cite{Mott_Sulfide_TMDC}, regarded to be emergent phenomena at low temperatures. However, it turns out to be rather challenging to realize strong correlations of electrons in the TMDC family. Nature of phase transitions seems to be determined by the band structure essentially, i.e., within the weak-coupling approach. The quasi two dimensional nature of the lattice structure does not cause sufficient anisotropy in the electronic structure except for several cases mentioned above.

\begin{figure}[h]
\includegraphics[width=0.4\textwidth]{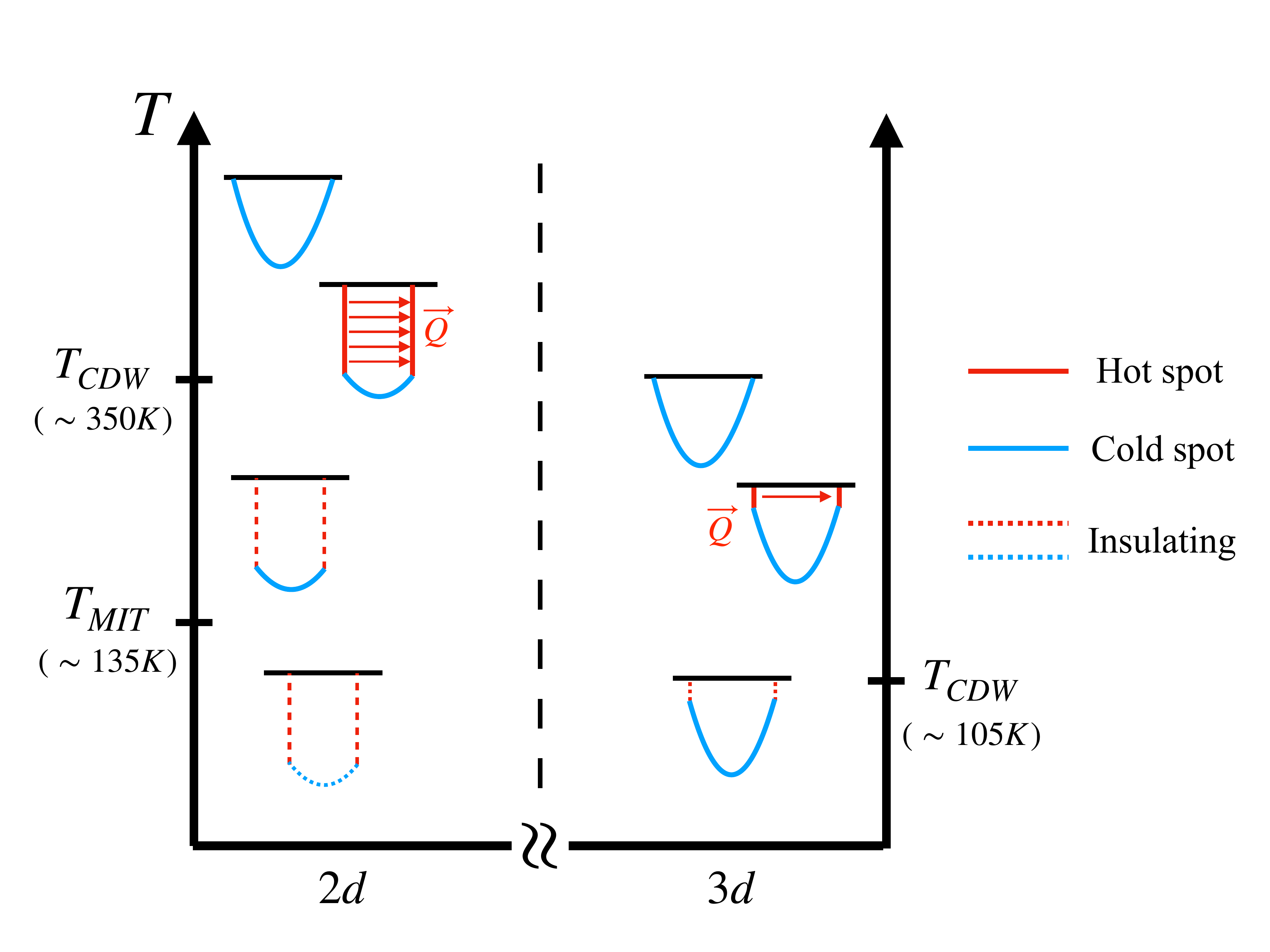}
\caption{Schematic phase diagram of monolayer $VSe_2$ and comparison with that of three-dimensional bulk $VSe_2$. Reducing thickness of $VSe_{2}$ from three dimensions to two dimensions, electron-electron correlations are enhanced to cause strong renormalization of the Fermi velocity. As a result, the weakly nested Fermi surface in three dimensions evolves into perfect Fermi surface nesting in two dimensions, observed in recent ARPES measurements \cite{ARPES_STM_CDW_MIT}. Another phase transition has been observed in monolayer $VSe_2$ \cite{ARPES_STM_CDW_MIT}, identified with a metal-insulator one. In this study we focus on the high-temperature CDW transition.}
\label{2D_Phase_VSe2.pdf}
\end{figure}

Recent measurements based on angle resolved photoemission spectroscopy (ARPES) and scanning tunneling microscopy (STM) \cite{ARPES_STM_CDW_MIT} have claimed that reducing thickness of three dimensional or quasi two dimensional TMDCs on appropriate substrates can give rise to drastic enhancement of correlation effects, responsible for novel nature of phase transitions. Such experiments found two types of charge ordering transitions driven by enhanced electron-electron and electron-phonon interactions in monolayer vanadium diselenide ($VSe_{2}$) on graphene substrates. It is well established that $VSe_{2}$ three dimensional bulk samples show a charge density wave (CDW) transition around $T_{c} \sim 105$ $K$, originating from Fermi surface nesting properties of ``hot" electrons \cite{CDW_VSe2_Bulk_I,CDW_VSe2_Bulk_II,CDW_VSe2_Bulk_III}. The low temperature CDW state coexists with ``cold" electrons of residual Fermi surfaces in this bulk system, keeping their metallicity. In comparison with this three dimensional case, ARPES measurements uncovered that the Fermi surface nesting becomes perfect in the monolayer limit, that is, the dynamics of such hot electrons is dispersionless along the orthogonal direction of the nesting wave-vector \cite{ARPES_STM_CDW_MIT}. This perfect Fermi-surface nesting property has been speculated to cause noticeable enhancement of the CDW transition temperature in two dimensions, even above the room temperature, implying the dynamics of strongly correlated electrons beyond the dynamics of three dimensional hot electrons. More interestingly, the ARPES experiment revealed that residual Fermi surfaces around cold zones disappear at $T_{c} \sim 135$ $K$, where an insulating phase is realized, never observed in the bulk system \cite{ARPES_STM_CDW_MIT}. STM measurements showed that this metal-insulator transition is driven by lattice distortions along a particular one dimensional direction \cite{ARPES_STM_CDW_MIT}. These experimental results are summarized in Fig. \ref{2D_Phase_VSe2.pdf}.

Nature of phase transitions in two dimensions turns out to differ from that in three dimensions, where the effective theory referred to as Hertz-Moriya-Millis theory \cite{HMM_EFT_QCP_I,HMM_EFT_QCP_II,HMM_EFT_QCP_III,HMM_EFT_QCP_IV}, regarded to be a mean-field theory, does not function. Even if electrons are weakly correlated at high temperatures, they become strongly correlated in the vicinity of two dimensional phase transitions involved with Fermi-surface instabilities \cite{QCP_Strong_Correlation}. This implies that the band structure itself can be renormalized rather drastically beyond the mean-field theoretical framework in the monolayered system.

In the present study we investigate how two dimensionality in dynamics of hot electrons affects the nature of the CDW transition in monolayer $VSe_{2}$. We construct an effective field theory in terms of hot electrons and critical CDW fluctuations. Recalling that the STM experiment confirmed that the CDW state shows essentially the same modulation pattern as the three dimensional system of $VSe_{2}$ \cite{ARPES_STM_CDW_MIT}, we assume imperfect Fermi surface nesting for hot electrons, where the nesting vector is given by the ``three dimensional" (quasi two dimensional) CDW ordering structure. Based on this effective field theory, we perform the renormalization group analysis in the scheme of a recently developed dimensional regularization for a Fermi-surface problem \cite{QCP_Dimensional_Regularization}. Our renormalization group analysis confirms that imperfect nesting universally flows into perfect nesting in two dimensions, where the Fermi velocity along the orthogonal direction of the nesting vector vanishes generically. We argue that this electronic reconstruction is responsible for the observation that the much higher CDW transition temperature $T_{2d,CDW} \sim 350K$ in two dimensional sample compared to the CDW transition temperature of three dimensional sample $T_{3d, CDW} \sim 105K$.

Before going further, we would like to speculate on the role of disorder in the high-temperature CDW and the low-temperature metal-insulator transitions of monolayer $VSe_{2}$ \cite{ARPES_STM_CDW_MIT}. It is well accepted that the diffusive dynamics of electrons gives rise to enhancement of electron-electron correlations, resulting from reinforcement of the interaction vertex and referred to as the Altshuer-Aronov correction \cite{Disorder_Review,AA_Correction}. One may expect that electron-phonon interaction vertices would be also amplified, responsible for both the drastic enhancement of the CDW critical temperature and the appearance of the metal-insulator transition in monolayer $VSe_{2}$. However, we point out that the CDW state shows essentially the same modulation pattern as the three dimensional system of $VSe_{2}$, revealed by the STM measurement \cite{ARPES_STM_CDW_MIT}, which implies that the CDW ordering results from the mechanism of Fermi surface nesting. In addition, it turns out that the CDW gap follows the Bardeen-Cooper-Schrieffer (BCS) type description well. These two experimental results suggest that the monolayer $VSe_{2}$ of Ref. \cite{ARPES_STM_CDW_MIT} does not lie in the strong disorder regime. Based on this discussion, we focus on the scenario of two dimensional CDW criticality in the present study.

\section{Effective field theory}

To construct an effective field theory, we recall the Fermi surface map of monolayer $VSe_2$, recently clarified in ARPES measurements \cite{ARPES_STM_CDW_MIT}. Compared with the electronic structure of three-dimensional bulk $VSe_2$ \cite{CDW_VSe2_Bulk_III}, the Fermi-surface structure of monolayer $VSe_2$ shows its qualitatively distinguished feature in the respect that the strong $k_{z}$ dispersion in three dimensions disappears to become almost dispersionless in two dimensions and the Fermi-surface nesting property is more enhanced enough to be called ``perfect" in monolayer $VSe_2$ when measured at $180$ $K$ \cite{ARPES_STM_CDW_MIT}. This ARPES experiment suggests a simplified Fermi-surface model for monolayer $VSe_2$, which shows a hexagonal Brillouin zone with six cigar-shaped electron pockets centered at the M points. See Fig. \ref{FermiSurface} for the two-dimensional Fermi-surface structure of monolayer $VSe_2$.

An important point in this simplified Fermi-surface model is that the Fermi-surface nesting is assumed to be weakly realized, which may sound to be contradictory with the emergence of perfect Fermi surface nesting in two dimensions. Moreover, this weak Fermi-surface nesting property has not been verified in recent ARPES measurements for monolayer $VSe_2$ \cite{ARPES_STM_CDW_MIT}. Based on the ARPES measurement and conventional BCS fitting for the CDW gap, the critical temperature for the CDW ordering is estimated (extrapolated) to be $T_{2d,CDW} \sim 350K$ for monolayer $VSe_2$, which lies above the measurement temperature \cite{ARPES_STM_CDW_MIT}. In other words, the electronic structure only below the critical temperature of the CDW metallic phase has been verified. Even if there exists a CDW gap in the Fermi surface structure, one can trace the nesting property experimentally, suggesting the emergence of perfect Fermi-surface nesting in monolayer $VSe_2$. In this study we assume a general Fermi surface structure, expected to appear much above the critical temperature $T_{2d,CDW} \sim 350K$ for monolayer $VSe_2$. Starting from this high-temperature Fermi-surface structure, we show the emergence of perfect Fermi surface nesting at low temperatures in two dimensions while the evolution of the Fermi surface structure with respect to temperature does not occur in three dimensions.

\def\b{4} 

\begin{figure}[h]
\centering

\begin{tikzpicture}[scale=.7]
\draw[style=thick] (\b,0)--(\b/2,{4*sqrt(3)/2})--(-\b/2,{\b*sqrt(3)/2})--(-\b,0)--(-\b/2,{-\b*sqrt(3)/2})--(\b/2,-{\b*sqrt(3)/2})--(4,0);

\draw[style=thin] (0.8,{\b*sqrt(3)/2}) arc (0:-90:0.8 and 2);
\draw[style=thin] (-0.8,{\b*sqrt(3)/2}) arc (-180:-90:0.8 and 2);
\draw[style=thin,rotate around={60:({-\b/2-1.2*cos(60)},{\b/2*sqrt(3)-1.2*sin(60)})}] ({-\b/2-1.2*cos(60)},{\b/2*sqrt(3)-1.2*sin(60)}) arc (0:-90:0.8 and 2);
\draw[style=thin,rotate around={60:({-\b/2-2.8*cos(60)},{\b/2*sqrt(3)-2.8*sin(60)})}] ({-\b/2-2.8*cos(60)},{\b/2*sqrt(3)-2.8*sin(60)}) arc(-180:-90:0.8 and 2);
\draw[style=thin,rotate around={-60:({\b/2+1.2*cos(60)},{\b/2*sqrt(3)-1.2*sin(60)})}] ({\b/2+1.2*cos(60)},{\b/2*sqrt(3)-1.2*sin(60)}) arc (-180:-90:0.8 and 2);
\draw[style=thin,rotate around={-60:({2+2.8*cos(60)},{2*sqrt(3)-2.8*sin(60)})}] ({\b/2+2.8*cos(60)},{\b/2*sqrt(3)-2.8*sin(60)}) arc(0:-90:0.8 and 2);

\path[fill=cyan, fill opacity=0.3] (0.8,{\b*sqrt(3)/2}) arc (0:-90:0.8 and 2)--(0,{\b*sqrt(3)/2})--(0.8,{\b*sqrt(3)/2});
\path[fill=cyan, fill opacity=0.3] (-0.8,{\b*sqrt(3)/2}) arc (-180:-90:0.8 and 2)--(0,{\b*sqrt(3)/2})--(-0.8,{\b*sqrt(3)/2});
\path[fill=cyan, fill opacity=0.3, rotate around={60:({-\b/2-1.2*cos(60)},{\b/2*sqrt(3)-1.2*sin(60)})}] ({-\b/2-1.2*cos(60)},{\b/2*sqrt(3)-1.2*sin(60)}) arc (0:-90:0.8 and 2)--({-\b/2-1.2*cos(60)-0.8},{\b/2*sqrt(3)-1.2*sin(60)})--({-\b/2-1.2*cos(60)},{\b/2*sqrt(3)-1.2*sin(60)});
\path[fill=cyan, fill opacity=0.3,rotate around={60:({-\b/2-2.8*cos(60)},{\b/2*sqrt(3)-2.8*sin(60)})}] ({-\b/2-2.8*cos(60)},{\b/2*sqrt(3)-2.8*sin(60)}) arc(-180:-90:0.8 and 2)--({-\b/2-2.8*cos(60)+0.8},{\b/2*sqrt(3)-2.8*sin(60)})--({-\b/2-2.8*cos(60)},{\b/2*sqrt(3)-2.8*sin(60)});
\path[fill=cyan, fill opacity=0.3, rotate around={-60:({\b/2+1.2*cos(60)},{\b/2*sqrt(3)-1.2*sin(60)})}] ({\b/2+1.2*cos(60)},{\b/2*sqrt(3)-1.2*sin(60)}) arc (-180:-90:0.8 and 2)--({\b/2+1.2*cos(60)+0.8},{\b/2*sqrt(3)-1.2*sin(60)})--({\b/2+1.2*cos(60)},{\b/2*sqrt(3)-1.2*sin(60)});
\path[fill=cyan, fill opacity=0.3, rotate around={-60:({2+2.8*cos(60)},{2*sqrt(3)-2.8*sin(60)})}] ({\b/2+2.8*cos(60)},{\b/2*sqrt(3)-2.8*sin(60)}) arc(0:-90:0.8 and 2)--({\b/2+2.8*cos(60)-0.8},{\b/2*sqrt(3)-2.8*sin(60)})--({\b/2+2.8*cos(60)},{\b/2*sqrt(3)-2.8*sin(60)});

\draw[style=thin] (0.8,-{\b*sqrt(3)/2}) arc (0:90:0.8 and 2);
\draw[style=thin] (-0.8,-{\b*sqrt(3)/2}) arc (180:90:0.8 and 2);
\draw[style=thin,rotate around={-60:({-\b/2-1.2*cos(60)},{-\b/2*sqrt(3)+1.2*sin(60)})}] ({-\b/2-1.2*cos(60)},{-\b/2*sqrt(3)+1.2*sin(60)}) arc (0:90:0.8 and 2);
\draw[style=thin,rotate around={-60:({-\b/2-2.8*cos(60)},{-\b/2*sqrt(3)+2.8*sin(60)})}] ({-\b/2-2.8*cos(60)},{-\b/2*sqrt(3)+2.8*sin(60)}) arc(180:90:0.8 and 2);
\draw[style=thin,rotate around={60:({\b/2+1.2*cos(60)},{-\b/2*sqrt(3)+1.2*sin(60)})}] ({\b/2+1.2*cos(60)},{-\b/2*sqrt(3)+1.2*sin(60)}) arc (180:90:0.8 and 2);
\draw[style=thin,rotate around={60:({\b/2+2.8*cos(60)},{-\b/2*sqrt(3)+2.8*sin(60)})}] ({\b/2+2.8*cos(60)},{-\b/2*sqrt(3)+2.8*sin(60)}) arc(0:90:0.8 and 2);

\path[fill=cyan, fill opacity=0.3] (0.8,-{\b*sqrt(3)/2}) arc (0:90:0.8 and 2)--(0,-{\b*sqrt(3)/2})--(0.8,-{\b*sqrt(3)/2});
\path[fill=cyan, fill opacity=0.3] (-0.8,-{\b*sqrt(3)/2}) arc (180:90:0.8 and 2)--(0,-{\b*sqrt(3)/2})--(-0.8,-{\b*sqrt(3)/2});
\path[fill=cyan, fill opacity=0.3,rotate around={-60:({-\b/2-1.2*cos(60)},{-\b/2*sqrt(3)+1.2*sin(60)})}] ({-\b/2-1.2*cos(60)},{-\b/2*sqrt(3)+1.2*sin(60)}) arc (0:90:0.8 and 2)--({-\b/2-1.2*cos(60)-0.8},{-\b/2*sqrt(3)+1.2*sin(60)})--({-\b/2-1.2*cos(60)},{-\b/2*sqrt(3)+1.2*sin(60)});
\path[fill=cyan, fill opacity=0.3, rotate around={-60:({-\b/2-2.8*cos(60)},{-\b/2*sqrt(3)+2.8*sin(60)})}] ({-\b/2-2.8*cos(60)},{-\b/2*sqrt(3)+2.8*sin(60)}) arc(180:90:0.8 and 2)--({-\b/2-2.8*cos(60)+0.8},{-\b/2*sqrt(3)+2.8*sin(60)})--({-\b/2-2.8*cos(60)},{-\b/2*sqrt(3)+2.8*sin(60)});
\path[fill=cyan, fill opacity=0.3, rotate around={60:({\b/2+1.2*cos(60)},{-\b/2*sqrt(3)+1.2*sin(60)})}] ({\b/2+1.2*cos(60)},{-\b/2*sqrt(3)+1.2*sin(60)}) arc (180:90:0.8 and 2)--({\b/2+1.2*cos(60)+0.8},{-\b/2*sqrt(3)+1.2*sin(60)})--({\b/2+1.2*cos(60)},{-\b/2*sqrt(3)+1.2*sin(60)});
\path[fill=cyan, fill opacity=0.3, rotate around={60:({\b/2+2.8*cos(60)},{-\b/2*sqrt(3)+2.8*sin(60)})}] ({\b/2+2.8*cos(60)},{-\b/2*sqrt(3)+2.8*sin(60)}) arc(0:90:0.8 and 2)--({\b/2+2.8*cos(60)-0.8},{-\b/2*sqrt(3)+2.8*sin(60)})--({\b/2+2.8*cos(60)},{-\b/2*sqrt(3)+2.8*sin(60)}) arc(0:90:0.8 and 2);

\draw[->] (0,0)--(1,0) node[pos=0.9,below] {$x$};
\draw[->](0,0)--(0,1) node[pos=0.9,left] {$y$};

\fill[color=blue] (0.8,{\b/2*sqrt(3)-0.3}) circle (0.1) node[right,black] {$1-$};
\fill[color=red] (-0.8,{\b/2*sqrt(3)-0.3}) circle (0.1) node[left,black] {$1+$};
\fill[color=red] (0.8,{-\b/2*sqrt(3)+0.3}) circle (0.1) node[right,black] {$\bar{1}+$};
\fill[color=blue] (-0.8,{-\b/2*sqrt(3)+0.3}) circle (0.1) node[left,black] {$\bar{1}-$};

\fill[color=blue] ({0.4-sqrt(3)*(\b/2*sqrt(3)-0.3)/2},{sqrt(3)*0.4+(\b/2*sqrt(3)-0.3)/2}) circle (0.1) node[right,black] {$2-$};
\fill[color=red] ({-0.4-sqrt(3)*(\b/2*sqrt(3)-0.3)/2},{-sqrt(3)*0.4+(\b/2*sqrt(3)-0.3)/2}) circle (0.1) node[below,black] {$2+$};
\fill[color=blue] ({-0.4+sqrt(3)*(\b/2*sqrt(3)-0.3)/2},{-sqrt(3)*0.4-(\b/2*sqrt(3)-0.3)/2}) circle (0.1) node[left,black] {$\bar{2}-$};
\fill[color=red] ({0.4+sqrt(3)*(\b/2*sqrt(3)-0.3)/2},{sqrt(3)*0.4-(\b/2*sqrt(3)-0.3)/2}) circle (0.1) node[above,black] {$\bar{2}+$};

\fill[color=blue] ({-0.4-sqrt(3)*(\b/2*sqrt(3)-0.3)/2},{sqrt(3)*0.4-(\b/2*sqrt(3)-0.3)/2}) circle (0.1) node[above,black] {$3-$};
\fill[color=red] ({0.4-sqrt(3)*(\b/2*sqrt(3)-0.3)/2},{-sqrt(3)*0.4-(\b/2*sqrt(3)-0.3)/2}) circle (0.1) node[right,black] {$3+$};
\fill[color=blue] ({0.4+sqrt(3)*(\b/2*sqrt(3)-0.3)/2},{-sqrt(3)*0.4+(\b/2*sqrt(3)-0.3)/2}) circle (0.1) node[below,black] {$\bar{3}-$};
\fill[color=red] ({-0.4+sqrt(3)*(\b/2*sqrt(3)-0.3)/2},{sqrt(3)*0.4+(\b/2*sqrt(3)-0.3)/2}) circle (0.1) node[left,black] {$\bar{3}+$};

\draw[thick,<-] (0.8,{\b/2*sqrt(3)-0.3})--(-0.8,{\b/2*sqrt(3)-0.3}) node[pos=0.5,below,black] {$Q_1$};
\draw[thick,<-] (0.8,{-\b/2*sqrt(3)+0.3})--(-0.8,{-\b/2*sqrt(3)+0.3}) node[pos=0.5,above,black] {$Q_1$};

\draw[thick,->] ({-0.4-sqrt(3)*(\b/2*sqrt(3)-0.3)/2},{-sqrt(3)*0.4+(\b/2*sqrt(3)-0.3)/2}) -- ({0.4-sqrt(3)*(\b/2*sqrt(3)-0.3)/2},{sqrt(3)*0.4+(\b/2*sqrt(3)-0.3)/2}) node[pos=0.5,right,black] {$Q_2$};
\draw[thick,<-] ({0.4+sqrt(3)*(\b/2*sqrt(3)-0.3)/2},{sqrt(3)*0.4-(\b/2*sqrt(3)-0.3)/2})--({-0.4+sqrt(3)*(\b/2*sqrt(3)-0.3)/2},{-sqrt(3)*0.4-(\b/2*sqrt(3)-0.3)/2}) node[pos=0.5,left,black]{$Q_2$} ;

\draw[thick,->] ({0.4-sqrt(3)*(\b/2*sqrt(3)-0.3)/2},{-sqrt(3)*0.4-(\b/2*sqrt(3)-0.3)/2}) -- ({-0.4-sqrt(3)*(\b/2*sqrt(3)-0.3)/2},{sqrt(3)*0.4-(\b/2*sqrt(3)-0.3)/2}) node[pos=0.5,right,black] {$Q_3$};
\draw[thick,->] ({0.4+sqrt(3)*(\b/2*sqrt(3)-0.3)/2},{-sqrt(3)*0.4+(\b/2*sqrt(3)-0.3)/2})--({-0.4+sqrt(3)*(\b/2*sqrt(3)-0.3)/2},{sqrt(3)*0.4+(\b/2*sqrt(3)-0.3)/2}) node[pos=0.5,left,black]{$Q_3$};

\draw[blue,->] (-3,{-sqrt(3)})--(3,{sqrt(3)}) node[pos=0.4,below,blue] {$\mathbf{b}_1$};
\draw[blue,->] (3,{-sqrt(3)})--(-3,{sqrt(3)}) node[pos=0.4,below,blue] {$\mathbf{b}_2$};

\end{tikzpicture}
\caption{Schematic diagram for the Fermi surface structure of monolayer $1 T$ $VSe_2$. Blue colored regions show electron pockets, and blue and red dots refer to "hot" spots, where $\mathbf{b}_1$ and $\mathbf{b}_2$ are two reciprocal lattice vectors. $\mathbf{Q}_1$, $\mathbf{Q}_2$, and $\mathbf{Q}_3$ are three different CDW nesting vectors.} \label{FermiSurface}
\end{figure}
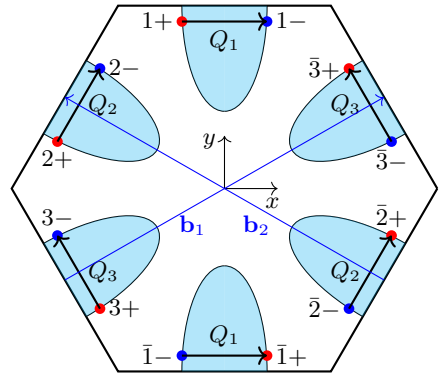

Based on this information, we construct an effective field theory as follows
\begin{widetext}
\begin{align}
S_0&=\sum_{n=1}^3\sum_{m=\pm}\int\frac{d^3k}{(2\pi)^3}\Big[\psi_n^{(m)*}(k)\Big(ik_0+\epsilon_n^{(m)}(k)\Big)\psi_n^{(m)}(k)+\psi_{\bar{n}}^{(m)*}(k)\Big(ik_0+\epsilon_{\bar{n}}^{(m)}(k)\Big)\psi_{\bar{n}}^{(m)}(k)\Big]\nonumber \\
&+\frac{1}{2}\sum_{n=1}^3\int\frac{d^3q}{(2\pi)^3}[q_0^2+c^2|\mathbf{q}|^2]\Phi_{Q_n}(q)\Phi_{Q_n}(-q) \\
S_{int-bf}&=e\sum_{n=1}^3\int \frac{d^3k}{(2\pi)^3}\int\frac{d^3q}{(2\pi)^3}\Phi_{Q_n}(q)\Big[\psi_n^{(-)*}(k+q)\psi_n^{(+)}(k)+\psi_n^{(+)*}(k+q)\psi_n^{(-)}(k)\nonumber \\
&+\psi_{\bar{n}}^{(+)*}(k+q)\psi_{\bar{n}}^{(-)}(k)+\psi_{\bar{n}}^{(-)*}(k+q)\psi_{\bar{n}}^{(+)}(k)\Big]\\
S_{int-b1}&=\frac{u_{1}}{4!}\sum_{n=1}^3\int \prod_{i=1}^4\frac{d^3q_i}{(2\pi)^3}\Phi_{Q_n}(q_1)\Phi_{Q_n}(q_2)\Phi_{Q_n}(q_3)\Phi_{Q_n}(q_4)(2\pi)^3\delta(q_1+q_2+q_3+q_4)\\
S_{int-b2}&=\frac{u_{2}}{2!2!}\int\prod_{i=1}^4\frac{d^3q_i}{(2\pi)^3}\Big[\Phi_{Q_1}(q_1)\Phi_{Q_1}(q_2)\Phi_{Q_2}(q_3)\Phi_{Q_2}(q_4)+\Phi_{Q_2}(q_1)\Phi_{Q_2}(q_2)\Phi_{Q_3}(q_3)\Phi_{Q_3}(q_4)\nonumber \\
&+\Phi_{Q_3}(q_1)\Phi_{Q_3}(q_2)\Phi_{Q_1}(q_3)\Phi_{Q_1}(q_4)\Big](2\pi)^3\delta(q_1+q_2+q_3+q_4) \\
S_{int-b3}&=\gamma\int \prod_{i=1}^3\frac{d^3q}{(2\pi)^3}\Big[\Phi_{Q_1}(q_1)\Phi_{Q_2}(q_2)\Phi_{Q_3}(q_3)+\Phi_{Q_1}(q_1)\Phi_{Q_2}(q_2)\Phi_{Q_3}(q_3)\Big] .
\end{align}
\end{widetext}
Here, $\psi_n^{(m)}(k)$ represents an electron field living in a hot spots denoted by $n ~ (\bar{n})=1,2,3 ~ (\bar{1},\bar{2},\bar{3})$ and $m=\pm$, as shown in Fig. \ref{FermiSurface}. These "hot" electrons are described by the dispersion relation $\epsilon_{n}^{(\pm)}(k)=\epsilon_1^{(\pm)}(R_{\theta_n}^{-1}k)=\pm k_{x,\theta_n}+vk_{y,\theta_n}$ and $\epsilon_{\bar{n}}^{(\pm)}(k)=-\epsilon_n^{(\pm)}(k)$, where $\Big(\begin{array}{c}k_{x,\theta_n} \\ k_{y,\theta_n}\end{array}\Big)=\Big(\begin{array}{cc} \cos\theta_n & \sin\theta_n \\ -\sin\theta_n & \cos\theta_n \end{array}\Big)\Big(\begin{array}{c} k_x \\ k_y \end{array}\Big)\equiv R_{\theta_n}^{-1}k$ with $\theta_n=\frac{\pi}{3}(n-1)$, which mimics the $C_{6}$ rotational symmetry of the Fermi-surface structure. It is clear that the Fermi-surface nesting becomes perfect when the velocity $v$ vanishes. $\Phi_{Q_n}(q)$ is a bosonic order parameter field to describe CDW fluctuations, where $Q_n$ with $n=1,2,3$ represents a CDW nesting wave vector of each Fermi surface. According to experiments of both two \cite{ARPES_STM_CDW_MIT} and three dimensional samples \cite{CDW_VSe2_Bulk_I}, the periodicity of the CDW order is four times of the lattice constant, given by $8 \vec{Q}_1 = \vec{b}_1 + \vec{b}_2$, which is commensurate. Generally, charge density fluctuations with a nesting vector $\mathbf{Q}$ can be described as follows
\begin{gather}
\rho(r)=e^{i\mathbf{Q}\cdot \mathbf{r}}\Phi_{Q}(r)+e^{-i\mathbf{Q}\cdot \mathbf{r}}\Phi_{Q}^{*}(r) ,
\end{gather}
where $\Phi_{-Q}(r)=\Phi^*_{Q}(r)$ has been used. Introducing $\Phi_{Q}(r)=\delta\rho(r)e^{i\theta(r)}$ into the above expression, we obtain $\rho(r)=\delta\rho(r)\cos(\mathbf{Q}\cdot\mathbf{r}+\theta(r))$. Here, $\delta\rho(r)$ and $\theta(r)$ represent amplitude and phase fluctuations of the CDW order parameter. In the case of commensurate CDW ordering, such phase fluctuations are irrelevant and neglected. As a result, we take into account $\Phi_{Q}(r)$ as a real valued function, i.e., $\Phi_{Q}(r)=\Phi^*_{Q}(r)$ \cite{CDW_OrderParameter}. These CDW fluctuations are assumed to follow the relativistic dispersion with their velocity $c$, regarded to be an effective field theory of the Ising model. Electrons connected by Fermi surface nesting are strongly correlated and described by $S_{int-bf}$ with an effective interaction parameter $e$. In addition, such CDW order parameters interact with themselves, constructed by symmetry consideration. $S_{int-b1}$ ($S_{int-b2}$) describes the self-interactions between CDW order parameters with the same momentum (different momenta) while $S_{int-b3}$ gives cubic self-interactions. Here, we do not take into account the $S_{int-b3}$ interaction for simplicity.

A conventional way solving this complex Fermi-surface problem is to take into account both self-energy corrections of electrons and order parameters self-consistently without considering vertex corrections, referred to as either Eliashberg theory or self-consistent random phase approximation \cite{HMM_EFT_QCP_I,HMM_EFT_QCP_II,HMM_EFT_QCP_III,HMM_EFT_QCP_IV}. This Fermi-surface problem has been regarded to be controlled in the so called large-N limit, where the spin degeneracy of electronic degrees of freedom is extended from $2$ to $N$ \cite{QCP_Large_N}. In other words, the Eliashberg theory is supposed to be exact in the $N\rightarrow\infty$ limit, where finite $N$ quantum corrections can be introduced in a controllable way, based on the solution of the Eliashberg theory. However, it turns out that this Fermi surface problem remains strongly correlated even in the $N\rightarrow\infty$ limit \cite{QCP_Strong_Correlation}, regarded to be a characteristic feature in two dimensions, which means that vertex corrections should be introduced self-consistently. Unfortunately, we do not know how to re-sum such quantum corrections consistently. Recently, the technique of ``graphenization" has been proposed as a way of controllable evaluation for Feynman diagrams, which generalizes the dimensional regularization technique for interacting boson problems into the Fermi surface problem, where the density of states is reduced to allow us to control effective interactions of electrons \cite{QCP_Dimensional_Regularization,Mott_Dimensional_Regularization}.

In order to prepare for the dimensional regularization scheme in the present problem, we introduce the two-component spinor
\begin{gather}
\Psi_n^{(\chi)}(k)=\Big(\begin{array}{c}\psi_n^{(\chi)}(k) \\ \chi\psi_{\bar{n}}^{(\chi)}(k)\end{array}\Big)
\end{gather}
and rewrite the above effective action in the following way
\begin{widetext}
\begin{align}
S_0&=\sum_{n=1}^{3}\sum_{m=\pm}\int\frac{d^3k}{(2\pi)^3}\Psi_n^{(m)\dagger}(k)[ik_0\tau^0+\epsilon_n^{(m)}(k)\tau^3]\Psi_n^{(m)}(k)\nonumber \\
&+\frac{1}{2}\sum_{n=1}^3\int\frac{d^3q}{(2\pi)^3}[q_0^2+c^2|\mathbf{q}|^2]\Phi_{n}(q)\Phi_{n}(-q)\\
S_{int-bf}&=e\sum_{n=1}^3 \int\frac{d^3k}{(2\pi)^3}\int\frac{d^3q}{(2\pi)^3} \Phi_n(q)\Bigg[\Psi_n^{(-)\dagger}(k+q)\tau^3\Psi_n^{(+)}(k)+\Psi_n^{(+)\dagger}(k+q)\tau^3\Psi_n^{(-)}(k)\Bigg]\\
S_{int-b1}&=\frac{u_{1}}{4!}\sum_{n=1}^3\int \prod_{i=1}^4\frac{d^3q_i}{(2\pi)^3}\Phi_{n}(q_1)\Phi_{n}(q_2)\Phi_{n}(q_3)\Phi_{n}(q_4)(2\pi)^3\delta(q_1+q_2+q_3+q_4)\\
S_{int-b2}&=\frac{u_{2}}{2!2!}\int\prod_{i=1}^4\frac{d^3q_i}{(2\pi)^3}\Big[\Phi_1(q_1)\Phi_1(q_2)\Phi_2(q_3)\Phi_2(q_4)+\Phi_2(q_1)\Phi_2(q_2)\Phi_3(q_3)\Phi_3(q_4)\nonumber \\
&+\Phi_3(q_1)\Phi_3(q_2)\Phi_1(q_3)\Phi_1(q_4)\Big](2\pi)^3\delta(q_1+q_2+q_3+q_4) ,
\end{align}
\end{widetext}
where $\tau^{3}$ is the Pauli matrix and $\Phi_{Q_n}(q)\equiv \Phi_{n}(q)$ in the short-hand notation.

Following S.-S. Lee's co-dimensional regularization method \cite{QCP_Dimensional_Regularization}, we write down the above two dimensional effective field theory in general $d$ dimensions
\begin{widetext}
\begin{align}
S_0&=\sum_{n=1}^3\sum_{m=\pm}\sum_{j=1}^{N_f}\int dk\bar{\Psi}_{n,j}^{(m)}(k)[i\mathbf{\Gamma}\cdot\mathbf{K}+i\gamma_{d-1}\epsilon_n^{(m)}(k_{d-1},k_d)]\Psi_{n,j}^{(m)}(k)\nonumber \\
&+\frac{1}{2}\sum_{n=1}^3\int dk [|\mathbf{K}|^2+c^2(k_{d-1}^2+k_d^2)]\Phi_n(k)\Phi_n(-k)\\
S_{int-bf}&=\frac{ie}{\sqrt{N_f}}\sum_{n=1}^{3}\sum_{j=1}^{N_f}\int dk\int dq \Phi_n(q)\Big[\bar{\Psi}_{n,j}^{(-)}(k+q)\gamma_{d-1}\Psi_{n,j}^{(+)}(k)+\bar{\Psi}_{n,j}^{(+)}(k+q)\gamma_{d-1}\Psi_{n,j}^{(-)}(k)\Big]\\
S_{int-b1}&=\frac{u_{1}}{4!}\sum_{n=1}^3\int \prod_{i=1}^4dq\Phi_{n}(q_1)\Phi_{n}(q_2)\Phi_{n}(q_3)\Phi_{n}(2\pi)^{d+1}\delta(q_1+q_2+q_3+q_4)\\
S_{int-b2}&=\frac{u_{2}}{2!2!}\int\prod_{i=1}^4dq_i\Big[\Phi_1(q_1)\Phi_1(q_2)\Phi_2(q_3)\Phi_2(q_4)+\Phi_2(q_1)\Phi_2(q_2)\Phi_3(q_3)\Phi_3(q_4)\nonumber \\
&+\Phi_3(q_1)\Phi_3(q_2)\Phi_1(q_3)\Phi_1(q_4)\Big](2\pi)^{d+1}\delta(q_1+q_2+q_3+q_4) .
\end{align}
\end{widetext}
Here, we increase the spatial dimension from $2$ to $d$. This procedure is encoded into the extension of momentum from $(k_0,k_x,k_y)$ to $(\mathbf{K},k_{d-1},k_d)$, where $\mathbf{K}=(k_0,k_1,\cdots ,k_{d-3},k_{d-2})\equiv (k_0,\mathbf{K}_\perp)$. Accordingly, the Dirac gamma matrix is changed from $(\gamma_0,\gamma_1,\gamma_2)$ to $(\mathbf{\Gamma},\gamma_{d-1},\gamma_d)$, where $\mathbf{\Gamma}=(\gamma_0,\gamma_1,\cdots,\gamma_{d-3},\gamma_{d-2})\equiv(\gamma_0,\mathbf{\Gamma}_\perp)$ with $\{\gamma_i,\gamma_j\}=2\delta_{ij}$. Although the number of components in the Dirac spinor should be enhanced to follow this dimensional generalization, we keep the nature of the two-component spinor. As shown below, it turns out that the upper critical dimension is $d = 3$, which enforces us to perform the renormalization group analysis slightly below this upper critical dimension. As a result, the two-component spinor is allowed. We also increase the number of fermion flavors from $1$ to $N_f$.

\section{Renormalization group analysis}

\subsection{Classical scaling}

It is straightforward to perform the scaling analysis in the effective action, resulting in
\begin{gather*}
\mathbf{K}=\frac{\mathbf{K}'}{b},\;\; k_{d-1}=\frac{k_{d-1}'}{b},\;\; k_{d}=\frac{k'_d}{b}\\
\Psi(k)=b^{\frac{d+2}{2}}\Psi'(k'),\;\; \Phi(k)=b^{\frac{d+3}{2}}\Phi(k'),\\
e=b^{\frac{d-3}{2}}e',\;\; u_{1}=b^{d-3}u_{1}',\;\; u_{2}=b^{d-3}u_{2}' .
\end{gather*}
Here, $b$ is the scaling parameter usually utilized in the Wilsonian scheme of the renormalization group analysis. It is related with $\mu$ as $b = \mu^{-1}$, where $\mu$ is the scaling parameter conventionally used in the high-energy physics scheme of the renormalization group analysis. See appendixes \ref{RGdetails} and \ref{1-loopCalculatoin}. As shown clearly in these equations, we observe that the upper critical dimension of all interaction parameters is $d_{c} = 3$, which allows us to perform the perturbative analysis in $d=3-\epsilon$ near the upper critical dimension, where $\epsilon$ is an expansion parameter.

\subsection{Renormalization group equations in the one-loop level}

We perform the renormalization group analysis based on the scheme usually utilized in high energy physics. We introduce an effective bare action in general $d-$dimensions. Introducing quantum corrections into this effective field theory, various ultraviolet (UV) divergences appear, but hidden in the intermediate stage, where the dimensional regularization is employed in this study. Such UV divergences are canceled by so called counterterms, where UV divergences are absorbed into some coefficients. Extracting all counterterms from the bare action, we have an effective renormalized action, where UV divergences disappear to be well defined. Then, it is straightforward to find relations between bare and renormalized quantities, where UV divergences are taken into account. Based on these relations, one can find renormalization group equations, referred to as $\beta-$functions, which describe how interaction parameters evolve from the high-temperature regime to the low-temperature region. Since this procedure is quite conventional, we would like to refer all details to appendixes \ref{RGdetails} and \ref{1-loopCalculatoin}.

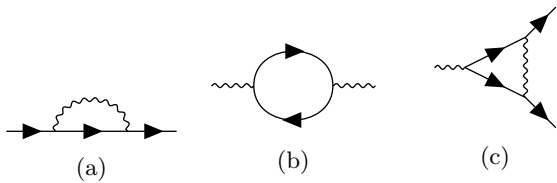
\begin{figure}[h]
\begin{subfigure}{0.13\textwidth}
\tikzsetnextfilename{FermiDiagram1}
\begin{tikzpicture}[baseline=1.8cm, scale=0.8]
\begin{feynhand}
\vertex (a) at (0,0); \vertex (b) at (0.8,0); \vertex (c) at (2,0); \vertex (d) at (2.8,0); \propagator[fermion] (a) to (b); \propagator[fermion] (b) to (c); \propagator[fermion] (c) to (d); \propagator[boson] (b) to [in=90,out=90,looseness=1.5](c);
\end{feynhand}
\end{tikzpicture}
\caption{}
\end{subfigure}
~
\begin{subfigure}{0.13\textwidth}
\tikzsetnextfilename{FermiDiagram2}
\begin{tikzpicture}[baseline=1.1cm,scale=0.8]
\begin{feynhand}
\vertex (a) at (0,0); \vertex (b) at (0.7,0); \vertex (c) at (2,0); \vertex (d) at (2.7,0); \propagator[boson] (a) to (b); \propagator[fermion] (b) to [in=90,out=90,looseness=1.5](c); \propagator[fermion] (c) to [in=-90,out=-90,looseness=1.5](b); \propagator[boson] (c) to (d);
\end{feynhand}
\end{tikzpicture}
\caption{}
\end{subfigure}
~
\begin{subfigure}{0.13\textwidth}
\tikzsetnextfilename{FermiDiagram3}
\begin{tikzpicture}[ scale=0.8]
\begin{feynhand}
\vertex (a) at (0,0); \vertex (b) at (0.5,0); \vertex (c) at (1.5,0.5); \vertex (d) at (2,1); \vertex (e) at (1.5,-0.5); \vertex (f) at (2,-1);
\propagator[boson] (a) to (b); \propagator[fermion] (b) to (c); \propagator[fermion] (c) to (d); \propagator[fermion] (b) to (e);
\propagator[fermion] (e) to (f); \propagator[boson] (c) to (e);
\end{feynhand}
\end{tikzpicture}
\caption{}
\end{subfigure}
\caption{One-loop diagrams for (a) Fermion self-energy, (b) Boson self-energy, and (c) Yukawa-type interaction vertex. Here, the solid (wavy) line represents the fermion (boson) propagator. See our Feynman rules in appendix \ref{RGdetails}.}
\label{1-loopDiagrams1}
\end{figure}

First, we consider the dynamical critical exponent $z$ and $\beta-$functions for the fermion velocity $v$, the boson (CDW order parameter) velocity $c$, and the effective coupling constant $e$ between electrons of hot spots and CDW fluctuations, given by
\begin{align}
z&=\underbrace{\frac{1}{1+\frac{e^2}{8\pi^2cN_f}[h_2(c,v)-h_1(c,v)]}\label{1loop_Beta1}}_{\text{Fig.}\ref{1-loopDiagrams1} (a)}\\
\beta_v&=\underbrace{\frac{ve^2zh_2(c,v)}{4\pi^2cN_f}}_{\text{Fig.}\ref{1-loopDiagrams1} (a)}\\
\beta_c&=\frac{e^2z}{16\pi^2N_f}\Big[\underbrace{2(h_2(c,v)-h_1(c,v))}_{\text{Fig.}\ref{1-loopDiagrams1}  (a)}+\underbrace{\frac{\pi cN_f}{2v}}_{\text{Fig.}\ref{1-loopDiagrams1}  (b)}\Big]\\
\beta_e&=\frac{ze}{2}\Big[-\epsilon+\underbrace{\frac{e^2}{16\pi v}}_{\text{Fig.}\ref{1-loopDiagrams1} (b)}+\frac{e^2}{4\pi^2cN_f}\Big(\underbrace{h_2(c,v)}_{\text{Fig.}\ref{1-loopDiagrams1}  (a)}\nonumber \\
&+\underbrace{\frac{1}{2}h_3(c,v)}_{\text{Fig.}\ref{1-loopDiagrams1}  (c)}\Big)\Big] . \label{YukawaBetaFunction}
\end{align}
We recall that the dynamical critical exponent tells us anisotropic scaling between space and time, related with the dispersion relation of critical CDW fluctuations or critical hot electrons. Here, quantum corrections to the dynamical critical exponent result from the fermion self-energy correction given by Fig. \ref{1-loopDiagrams1} (a), which leads the dynamical critical exponent to be larger than one, consistent with the causality condition of any local field theories. We used the short-hand notation for
\begin{gather}
h_1(c,v)=\int_0^1dx\sqrt{\frac{x}{xc^2+(1-x)(1+v^2)}} , \\ h_2(c,v)=\int_0^1 dx c^2\sqrt{\frac{x}{[xc^2+(1-x)(1+v^2)]^3}} ,
\end{gather}
both of which are positive definite. The renormalization group flow of the fermion velocity is given by the fermion self-energy correction [Fig. \ref{1-loopDiagrams1} (a)], where the fermion velocity decreases to vanish in the low-temperature limit. The boson velocity renormalization occurs from the boson self-energy correction [Fig. \ref{1-loopDiagrams1} (b)] while the fermion self-energy correction [Fig. \ref{1-loopDiagrams1} (a)] also contributes to the boson velocity renormalization, originating from the spacetime anisotropic scaling described by the dynamical critical exponent. It turns out that the space-time anisotropic scaling, if combined with the anomalous scaling dimension of the boson field given by the boson wave-function renormalization constant, enhances the boson velocity while the boson self-energy correction reduces it. The coupling constant for the Yukawa-type interaction vertex is renormalized by not only fermion [Fig. \ref{1-loopDiagrams1} (a)] and boson self-energy corrections [Fig. \ref{1-loopDiagrams1} (b)] but also vertex corrections [Fig. \ref{1-loopDiagrams1} (c)], where both self-energy corrections appear as anomalous scaling dimensions of fields, given by each wave-function renormalization constant. In addition, the anisotropic scaling between space and time also contributes. Here, the boson-fermion vertex function is given by
\begin{widetext}
\begin{gather}
h_3(c,v)=c\int_0^1dx\int_0^{1-x}dy \frac{2g_1(c,v,x,y)g_2(c,v,x,y)-2v^2(x-y)^2+g_2(c,v,x,y)-v^2g_1(c,v,x,y)}{[g_1(c,v,x,y)g_2(c,v,x,y)-v^2(x-y)^2]^{3/2}} , \\
g_1(c,v,x,y)=c^2(1-x-y)+x+y,\;\;\; g_2(c,v,x,y)=c^2(1-x-y)+v^2(x+y) .
\end{gather}
\end{widetext}
All quantum fluctuations, given by Fig. \ref{1-loopDiagrams1} and appropriately combined, screen out the boson-fermion effective interaction, thus reduced. We refer explicit calculations for frequency and momentum integrals given by the Feynman diagram Fig. \ref{1-loopDiagrams1} to appendix \ref{1-loopCalculatoin}.

\begin{figure}[h]
\begin{subfigure}[b]{0.13\textwidth}
\begin{tikzpicture}[scale=0.5]
\begin{feynhand}
\vertex (a) at (-1,1) {$n$}; \vertex (b) at (-1,-1) {$n$}; \vertex[ringdot] (c) at (0,0) {}; \vertex[ringdot] (d) at (1,0) {}; \vertex (e) at (2,1) {$n$}; \vertex (f) at (2,-1) {$n$};
\propagator[boson] (a) to (c); \propagator[boson] (b) to (c); \propagator[boson] (c) to [in=90,out=90,looseness=1.5](d); \propagator[boson] (d) to [in=-90,out=-90,looseness=1.5](c); \propagator[boson] (d) to (e); \propagator[boson] (d) to (f);
\end{feynhand}
\end{tikzpicture}
\caption{}
\end{subfigure}
~
\begin{subfigure}[b]{0.13\textwidth}
\begin{tikzpicture}[scale=0.5]
\begin{feynhand}
\vertex (a) at (-1,1) {$n$}; \vertex (b) at (-1,-1) {$n$}; \vertex[ringdot] (c) at (0,0) {}; \vertex[ringdot] (d) at (1,0) {}; \vertex (e) at (2,1) {$n$}; \vertex (f) at (2,-1) {$n$};
\propagator[boson] (a) to [in=30,out=0](d); \propagator[boson] (b) to (c); \propagator[boson] (c) to [in=90,out=90,looseness=1.5](d); \propagator[boson] (d) to [in=-90,out=-90,looseness=1.5](c); \propagator[boson] (c) to [in=180,out=120](e); \propagator[boson] (d) to (f);
\end{feynhand}
\end{tikzpicture}
\caption{}
\end{subfigure}
~
\begin{subfigure}[b]{0.13\textwidth}
\begin{tikzpicture}[scale=0.5]
\begin{feynhand}
\vertex (a) at (-1,1) {$n$}; \vertex (b) at (1,1) {$n$}; \vertex[ringdot] (c) at (0,0) {}; \vertex[ringdot] (d) at (0,-1) {}; \vertex (e) at (-1,-2) {$n$}; \vertex (f) at (1,-2) {$n$};
\propagator[boson] (a) to (c); \propagator[boson] (b) to (c); \propagator[boson] (c) to [in=0,out=0,looseness=1.5](d); \propagator[boson] (d) to [in=180,out=180,looseness=1.5](c); \propagator[boson] (d) to (e); \propagator[boson] (d) to (f);
\end{feynhand}
\end{tikzpicture}
\caption{}
\end{subfigure}
\\
\begin{subfigure}[b]{0.13\textwidth}
\begin{tikzpicture}[scale=0.5]
\begin{feynhand}
\vertex (a) at (-1,1) {$n$}; \vertex (b) at (-1,-1) {$n$}; \vertex[dot] (c) at (0,0) {}; \vertex[dot] (d) at (1,0) {}; \vertex (e) at (2,1) {$n$}; \vertex (f) at (2,-1) {$n$};
\propagator[boson] (a) to (c); \propagator[boson] (b) to (c); \propagator[boson] (c) to [in=90,out=90,looseness=1.5](d); \propagator[boson] (d) to [in=-90,out=-90,looseness=1.5](c); \propagator[boson] (d) to (e); \propagator[boson] (d) to (f);
\end{feynhand}
\end{tikzpicture}
\caption{}
\end{subfigure}
~
\begin{subfigure}[b]{0.13\textwidth}
\begin{tikzpicture}[scale=0.5]
\begin{feynhand}
\vertex (a) at (-1,1) {$n$}; \vertex (b) at (-1,-1) {$n$}; \vertex[dot] (c) at (0,0) {}; \vertex[dot] (d) at (1,0) {}; \vertex (e) at (2,1) {$n$}; \vertex (f) at (2,-1) {$n$};
\propagator[boson] (a) to [in=30,out=0](d); \propagator[boson] (b) to (c); \propagator[boson] (c) to [in=90,out=90,looseness=1.5](d); \propagator[boson] (d) to [in=-90,out=-90,looseness=1.5](c); \propagator[boson] (c) to [in=180,out=120](e); \propagator[boson] (d) to (f);
\end{feynhand}
\end{tikzpicture}
\caption{}
\end{subfigure}
~
\begin{subfigure}[b]{0.13\textwidth}
\begin{tikzpicture}[scale=0.5]
\begin{feynhand}
\vertex (a) at (-1,1) {$n$}; \vertex (b) at (1,1) {$n$}; \vertex[dot] (c) at (0,0) {}; \vertex[dot] (d) at (0,-1) {}; \vertex (e) at (-1,-2) {$n$}; \vertex (f) at (1,-2) {$n$};
\propagator[boson] (a) to (c); \propagator[boson] (b) to (c); \propagator[boson] (c) to [in=0,out=0,looseness=1.5](d); \propagator[boson] (d) to [in=180,out=180,looseness=1.5](c); \propagator[boson] (d) to (e); \propagator[boson] (d) to (f);
\end{feynhand}
\end{tikzpicture}
\caption{}
\end{subfigure}
\caption{All one-loop diagrams for the $u_1$ boson interaction. The black (white) dot represents the $u_1$ ($u_{2}$) boson interaction vertex. See our Feynman rules in appendix \ref{RGdetails}.}
\label{1-loopDiagrams2}
\end{figure}
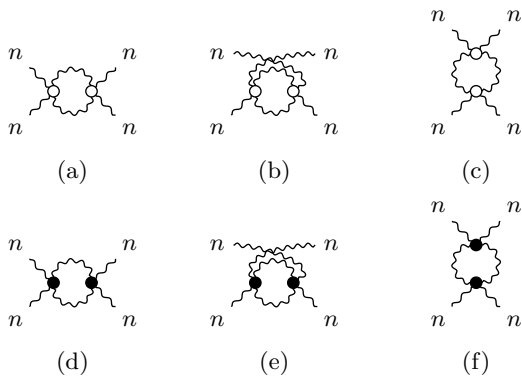

\begin{figure}[h]
\begin{subfigure}[b]{0.15\textwidth}
\begin{tikzpicture}[scale=0.5]
\begin{feynhand}
\vertex (a) at (-1,1) {$n$}; \vertex (b) at (-1,-1) {$n$}; \vertex[ringdot] (c) at (0,0) {}; \vertex[dot] (d) at (1,0) {}; \vertex (e) at (2,1) {$\bar{n}$}; \vertex (f) at (2,-1) {$\bar{n}$};
\propagator[boson] (a) to (c); \propagator[boson] (b) to (c); \propagator[boson] (c) to [in=90,out=90,looseness=1.5](d); \propagator[boson] (d) to [in=-90,out=-90,looseness=1.5](c); \propagator[boson] (d) to (e); \propagator[boson] (d) to (f);
\end{feynhand}
\end{tikzpicture}
\caption{}
\end{subfigure}
~
\begin{subfigure}[b]{0.15\textwidth}
\begin{tikzpicture}[scale=0.5]
\begin{feynhand}
\vertex (a) at (-1,1) {$n$}; \vertex (b) at (-1,-1) {$n$}; \vertex[dot] (c) at (0,0) {}; \vertex[ringdot] (d) at (1,0) {}; \vertex (e) at (2,1) {$\bar{n}$}; \vertex (f) at (2,-1) {$\bar{n}$};
\propagator[boson] (a) to (c); \propagator[boson] (b) to (c); \propagator[boson] (c) to [in=90,out=90,looseness=1.5](d); \propagator[boson] (d) to [in=-90,out=-90,looseness=1.5](c); \propagator[boson] (d) to (e); \propagator[boson] (d) to (f);
\end{feynhand}
\end{tikzpicture}
\caption{}
\end{subfigure}
\\
\begin{subfigure}[b]{0.13\textwidth}
\begin{tikzpicture}[scale=0.5]
\begin{feynhand}
\vertex (a) at (-1,1) {$n$}; \vertex (b) at (-1,-1) {$n$}; \vertex[dot] (c) at (0,0) {}; \vertex[dot] (d) at (1,0) {}; \vertex (e) at (2,1) {$\bar{n}$}; \vertex (f) at (2,-1) {$\bar{n}$};
\propagator[boson] (a) to (c); \propagator[boson] (b) to (c); \propagator[boson] (c) to [in=90,out=90,looseness=1.5](d); \propagator[boson] (d) to [in=-90,out=-90,looseness=1.5](c); \propagator[boson] (d) to (e); \propagator[boson] (d) to (f);
\end{feynhand}
\end{tikzpicture}
\caption{}
\end{subfigure}
~
\begin{subfigure}[b]{0.13\textwidth}
\begin{tikzpicture}[scale=0.5]
\begin{feynhand}
\vertex (a) at (-1,1) {$n$}; \vertex (b) at (-1,-1) {$n$}; \vertex[dot] (c) at (0,0) {}; \vertex[dot] (d) at (1,0) {}; \vertex (e) at (2,1) {$\bar{n}$}; \vertex (f) at (2,-1) {$\bar{n}$};
\propagator[boson] (a) to [in=30,out=0](d); \propagator[boson] (b) to (c); \propagator[boson] (c) to [in=90,out=90,looseness=1.5](d); \propagator[boson] (d) to [in=-90,out=-90,looseness=1.5](c); \propagator[boson] (c) to [in=180,out=120](e); \propagator[boson] (d) to (f);
\end{feynhand}
\end{tikzpicture}
\caption{}
\end{subfigure}
~
\begin{subfigure}[b]{0.13\textwidth}
\begin{tikzpicture}[scale=0.5]
\begin{feynhand}
\vertex (a) at (-1,1) {$n$}; \vertex (b) at (1,1) {$\bar{n}$}; \vertex[dot] (c) at (0,0) {}; \vertex[dot] (d) at (0,-1) {}; \vertex (e) at (-1,-2) {$n$}; \vertex (f) at (1,-2) {$\bar{n}$};
\propagator[boson] (a) to (c); \propagator[boson] (b) to (c); \propagator[boson] (c) to [in=0,out=0,looseness=1.5](d); \propagator[boson] (d) to [in=180,out=180,looseness=1.5](c); \propagator[boson] (d) to (e); \propagator[boson] (d) to (f);
\end{feynhand}
\end{tikzpicture}
\caption{}
\end{subfigure}
\caption{All one-loop diagrams for the $u_2$ boson interaction. One may be confused with diagrams (c), (d), and (e). We recall $n = 1,~2,~3$ for electron pockets. As a result, if both external lines are given by $n = 1$ and $\bar{n} = 3$, respectively, for example, the intermediate propagator line is indexed with $\tilde{n} = 2$.}
\label{1-loopDiagrams3}
\end{figure}
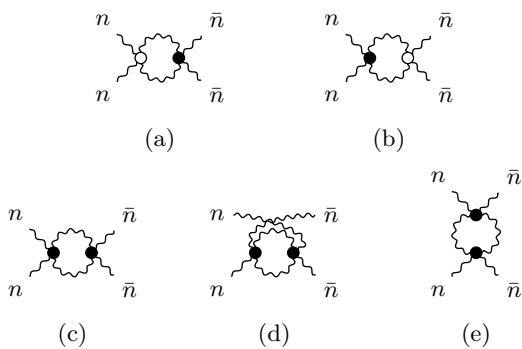

Second, we consider renormalizations of both boson interactions denoted by $u_1$ and $u_2$, shown in Fig. \ref{1-loopDiagrams2} for $u_1$ and \ref{1-loopDiagrams3} for $u_2$. We recall that $u_1$ is the self-interaction strength of CDW fluctuations with the same nesting momentum $Q_{n}$ while $u_2$ is that between $Q_{n}$ and $Q_{\bar{n}}$ bosons. One-loop beta functions for $u_1$ and $u_2$ are given as follows
\begin{align}
\beta_{u_1}&=zu_1\Big[-\epsilon+\overbrace{\frac{e^2}{4\pi^2 cN_f}[h_2(c,v)-h_1(c,v)]}^{\text{Fig.}\ref{1-loopDiagrams1} (a)}\nonumber \\
&+\underbrace{\frac{e^2}{8\pi v}}_{\text{Fig.}\ref{1-loopDiagrams1} (b)}+\underbrace{\frac{3u_1}{16\pi^2 c^2}}_{\text{\text{Fig.}\ref{1-loopDiagrams2}(a),(b),(c)}}+\underbrace{\frac{3u_2^2}{8\pi^2c^2u_1}}_{\text{Fig.}\ref{1-loopDiagrams2}(d),(e),(f)}\Big]\\
\beta_{u_2}&=zu_2\Big[-\epsilon+\underbrace{\frac{e^2}{4\pi^2 cN_f}[h_2(c,v)-h_1(c,v)]}_{\text{Fig.}\ref{1-loopDiagrams1} (a)}\nonumber \\
&+\underbrace{\frac{e^2}{8\pi v}}_{\text{Fig.}\ref{1-loopDiagrams1}(b)}+\underbrace{\frac{u_1}{8\pi^2 c^2}}_{\text{Fig.}\ref{1-loopDiagrams3}(a),(b)}+\underbrace{\frac{5u_2}{16\pi^2c^2}}_{\text{Fig.}\ref{1-loopDiagrams3}(c),(d),(e)}\Big]\label{1loop_Beta6}.
\end{align}
%
%
The renormalization group flow of the $u_1$ boson interaction is governed by the anisotropic scaling of the spacetime involved with the dynamical critical exponent [Fig. \ref{1-loopDiagrams1} (a)], the anomalous scaling exponent of the boson field given by the wave-function renormalization constant [Fig. \ref{1-loopDiagrams1} (b)], and renormalizations of the $u_1$ interaction vertex resulting from both $u_1$ [Fig. \ref{1-loopDiagrams2} (a), (b), (c)] and $u_2$ [Fig. \ref{1-loopDiagrams2} (d), (e), (f)] effective interactions. It turns out that all types of quantum corrections give rise to screening effects to the $u_1$ interaction except for the fact that the effect of the spacetime anisotropic scaling, if combined with the anomalous scaling dimension of the boson field, enhances the interaction strength. We recall our convention that beta functions with positive values mean that the corresponding coupling constant decreases as approaching to the low energy limit. The evolution behavior of the $u_2$ boson interaction is quite similar to that of $u_1$ except for the fact that screening effects are reduced slightly, compared with the $u_1$ case.
\\
\subsubsection{Beta function in the absence of the fermion-boson coupling}

\begin{figure}[h]
\includegraphics[width=0.3\textwidth]{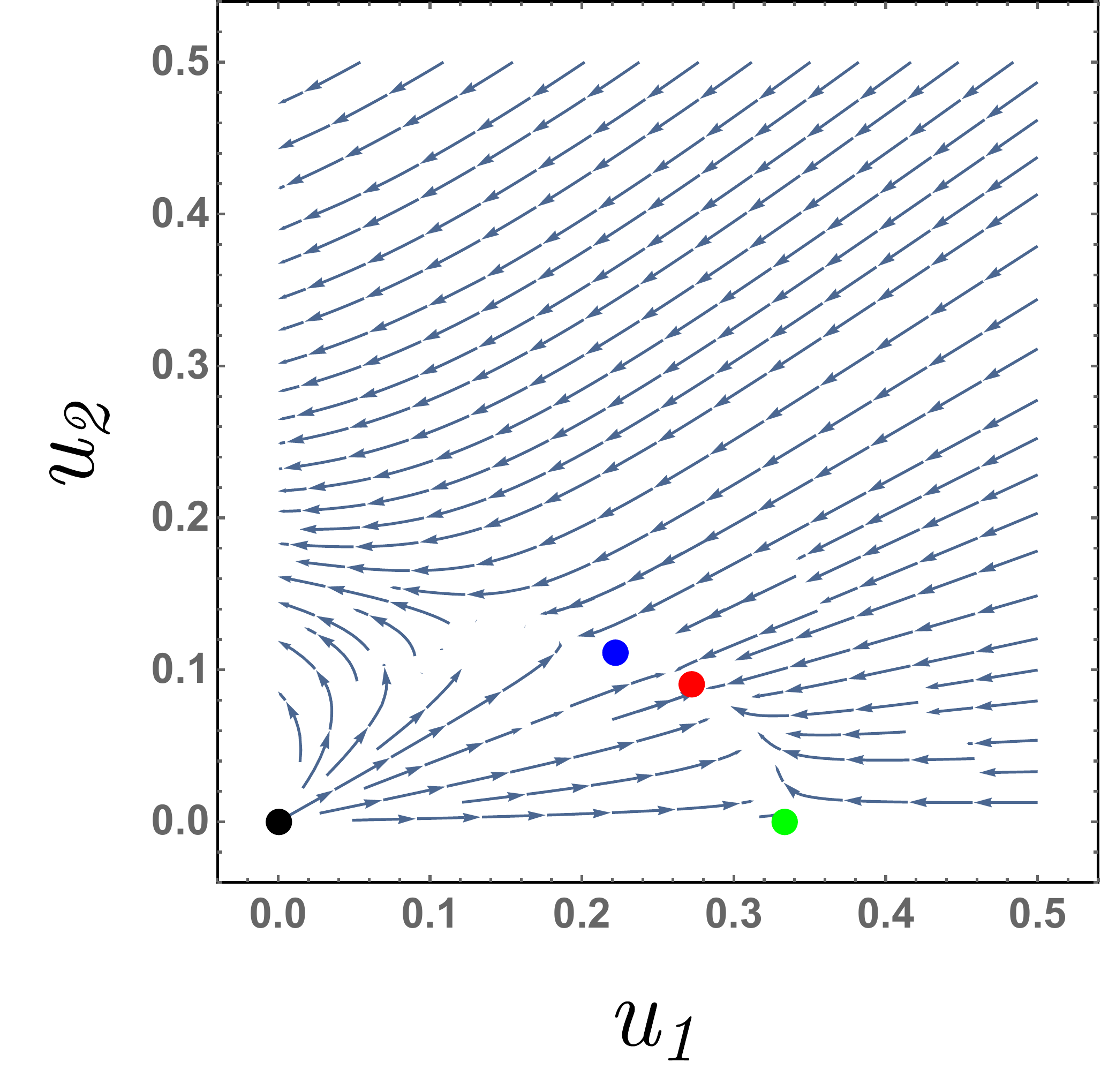}
\caption{Renormalization group flow diagram of $u_1$ and $u_2$ with $N_f=1$ and $\epsilon=0.01$ when $e=0$. The gaussian fixed point $(u_1, u_2) = (0, 0)$ denoted by the black dot is unstable against the presence of weak repulsive $u_1$ interactions, showing the renormalization group flow toward the conventional Wilson-Fisher fixed point $(16\pi^2c_0^2\epsilon/3,0)$ represented by the green dot. This Wilson-Fisher fixed point is destabilized by weak repulsive $u_2$ interactions, falling into a modified Wilson-Fisher fixed point $(48\pi^2c_0^2\epsilon/11,16\pi^2c_0^2/11)$ given by the red dot. There exists a line of separation, the fixed point on which is given by $(32\pi^2c_0^2\epsilon/0,16\pi^2c_0^2\epsilon/9)$ and marked by the blue dot: A runaway flow is observed toward a negative value of the boson interaction $u_1$ above this line while the renormalization group flow arrives at the modified Wilson-Fisher fixed point below.} \label{u1_u2_flow}
\end{figure}

Although it is not difficult to solve these coupled renormalization group equations, we start from the case in the absence of the fermion-boson interaction vertex, i.e., $e = 0$. Then, we focus on the renormalization group equations for $u_1$ and $u_2$ with $e = 0$. As a result, we find four fixed points shown in Fig. \ref{u1_u2_flow}. The gaussian fixed point $(u_1, u_2) = (0, 0)$, denoted by the black dot, becomes unstable to show the renormalization group flow to the conventional Wilson-Fisher fixed point $(16\pi^2c_0^2\epsilon/3,0)$, represented by the green dot \cite{RG_Equation_Corr}. This Wilson-Fisher fixed point is destabilized against the presence of weak $u_2$ interactions, falling into a modified Wilson-Fisher fixed point $(48\pi^2c_0^2\epsilon/11,16\pi^2c_0^2/11)$, given by the red dot. This $u_2$ modified Wilson-Fisher fixed point can be regarded as a critical point in the present continuous CDW transition. On the other hand, we find a line of separation, in the above of which a runaway flow is observed toward a negative value of the boson interaction $u_1$, but in the below of which the renormalization group flow arrives at the modified Wilson-Fisher fixed point. The fixed point on this line of separation is $(32\pi^2c_0^2\epsilon/0,16\pi^2c_0^2\epsilon/9)$, marked by the blue dot. The runaway flow toward a negative value of the $u_1$ interaction leads us to identify this fixed point with the parameter point for the fluctuation-induced first-order phase transition \cite{Fluctuation_Induced_First_Order_I,Fluctuation_Induced_First_Order_II,
Fluctuation_Induced_First_Order_III,Fluctuation_Induced_First_Order_IV}. This may result from interactions between competing CDW fluctuations with several ordering wave vectors.

\subsubsection{Beta function in the presence of the fermion-boson coupling}

\begin{widetext}

\begin{figure}[h]
\begin{subfigure}[b]{0.32\textwidth}
\includegraphics[width=\textwidth]{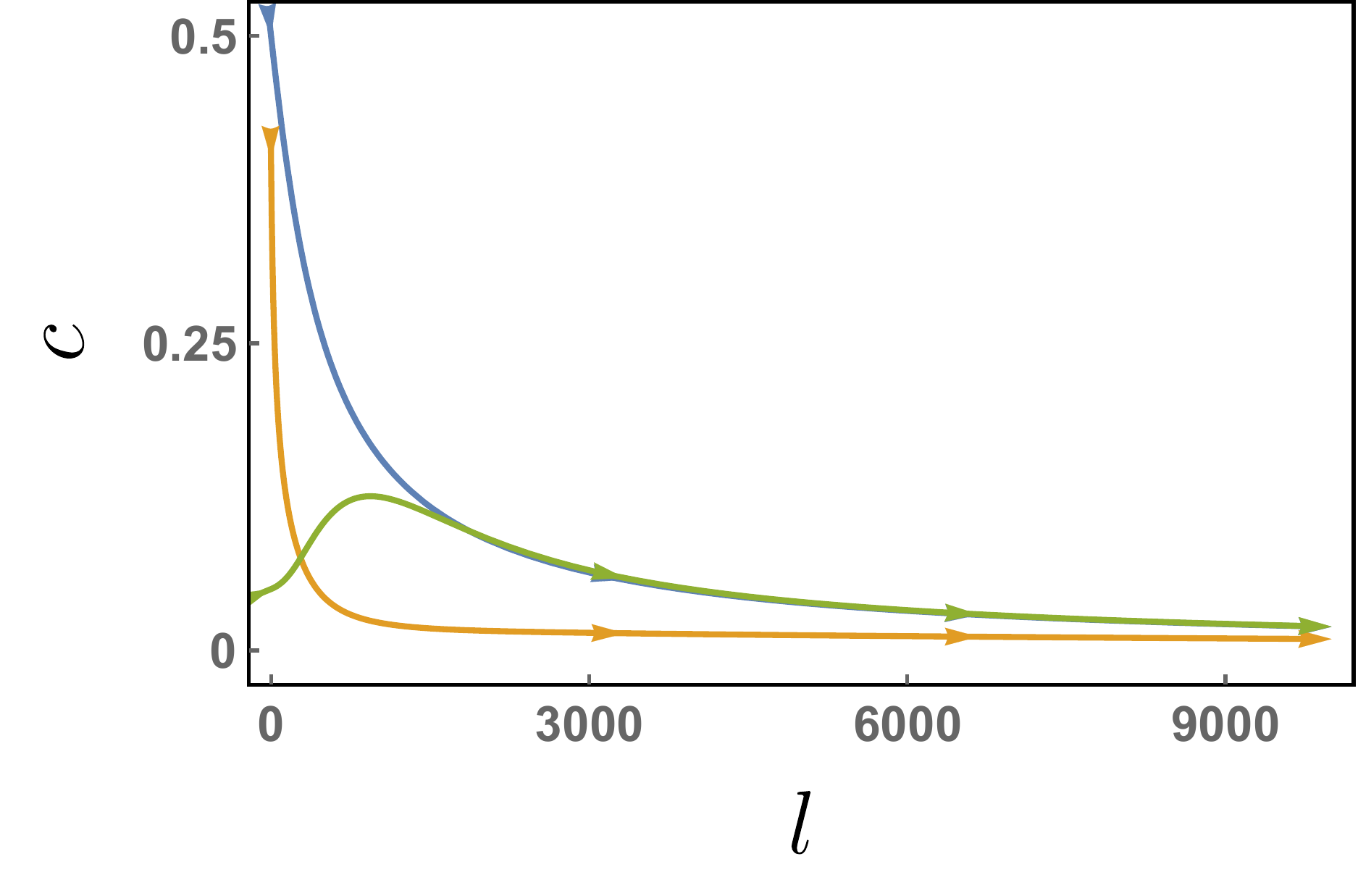}
\caption{RG flow of $c$}
\end{subfigure}
~
\begin{subfigure}[b]{0.32\textwidth}
\includegraphics[width=\textwidth]{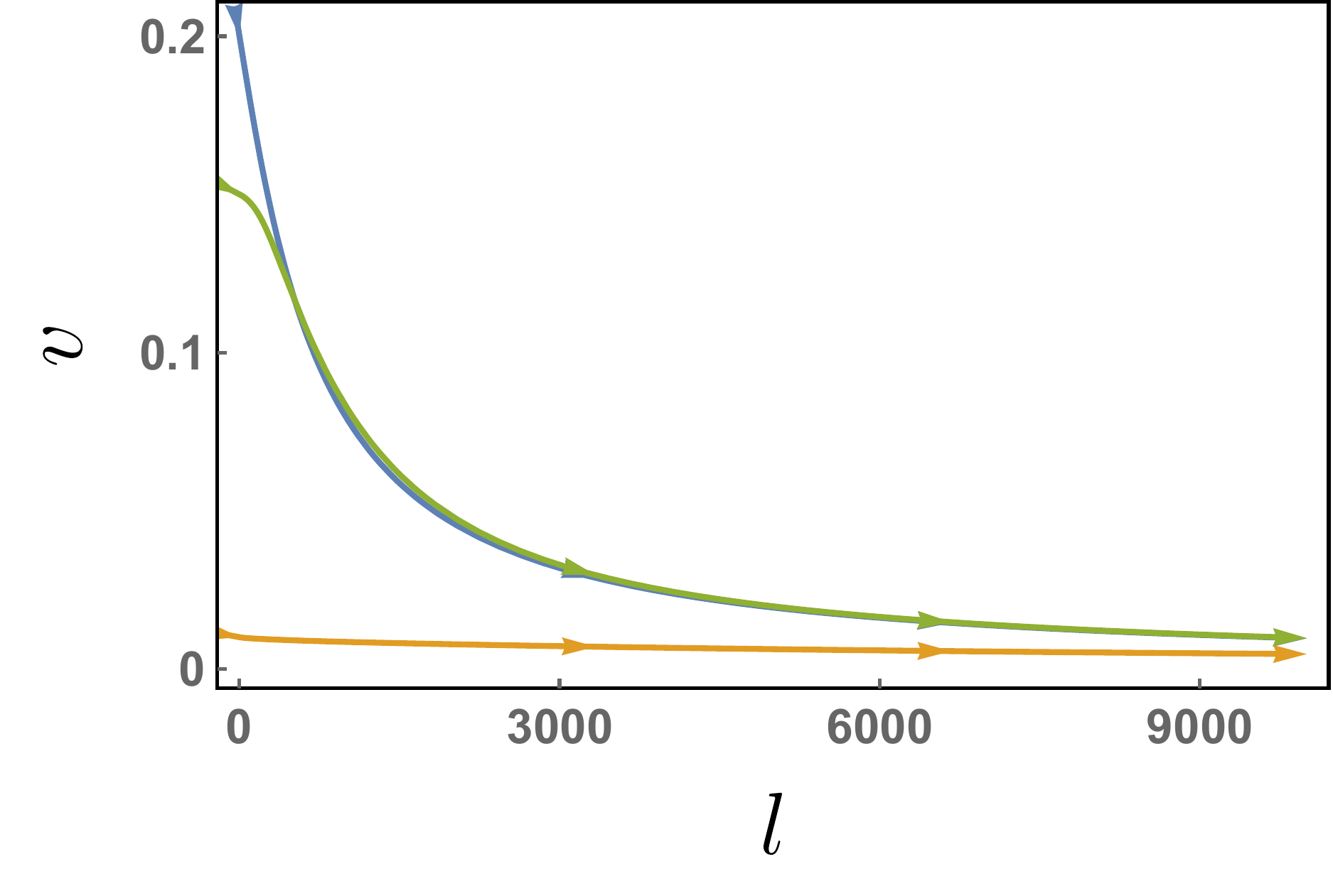}
\caption{RG flow of $v$}
\end{subfigure}
~
\begin{subfigure}[b]{0.32\textwidth}
\includegraphics[width=\textwidth]{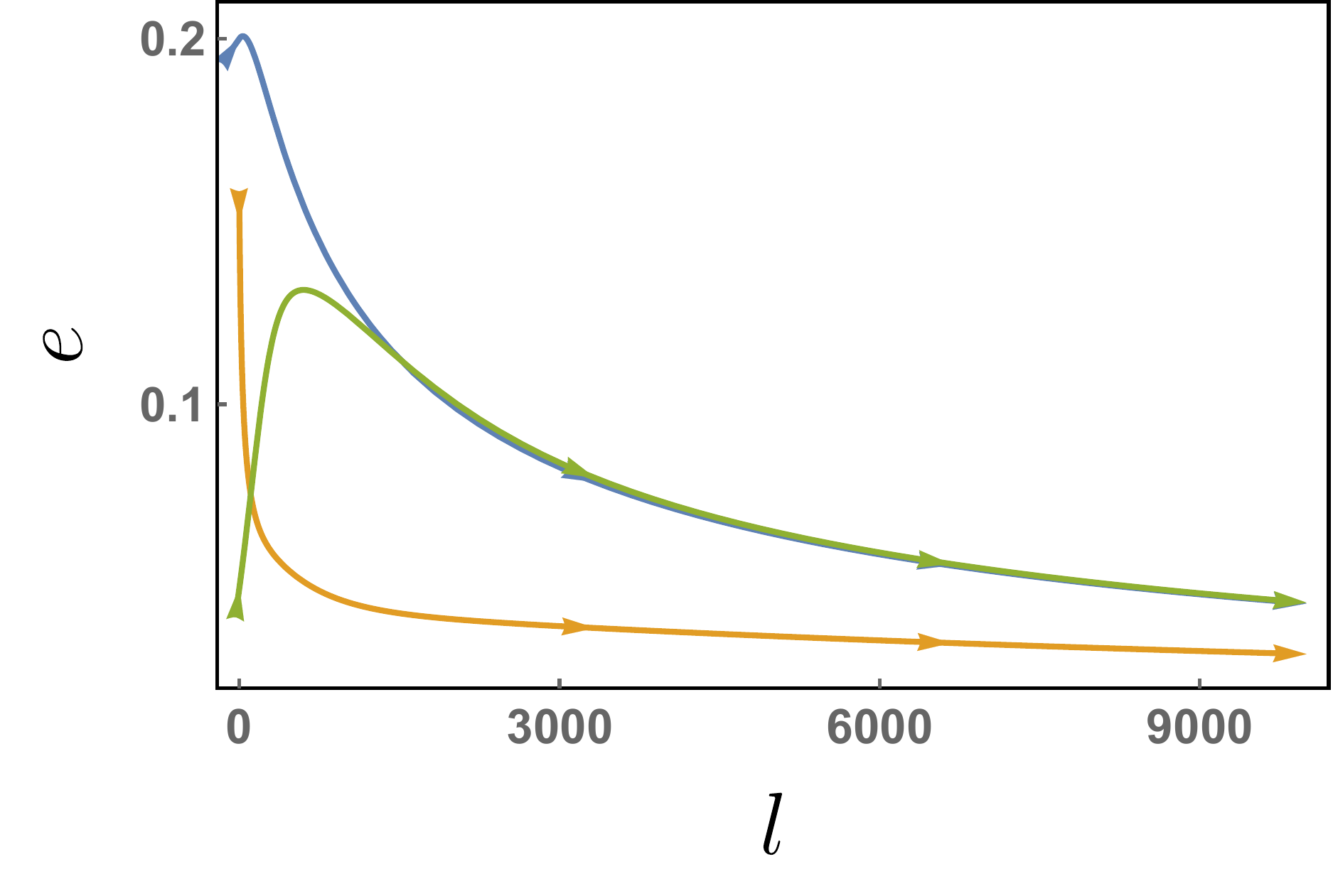}
\caption{RG flow of $e$}
\end{subfigure}
\caption{Renormalization group flow of $c$, $v$, and $e$ with $N_f=1$ and $\epsilon=0.01$ for different initial conditions, which correspond to $(c_0,v_0,e_0)=(0.5,0.2,0.2)$ for blue, (0.4,0.01,0.15) for orange, and (0.05,0.15,0.05) for green, respectively. $l$ is $-\ln \mu$, which increases as energy is lowered.}
\label{c_v_e_betafunction_plot}
\end{figure}

\end{widetext}

\begin{figure}[h]
\includegraphics[width=0.25\textwidth]{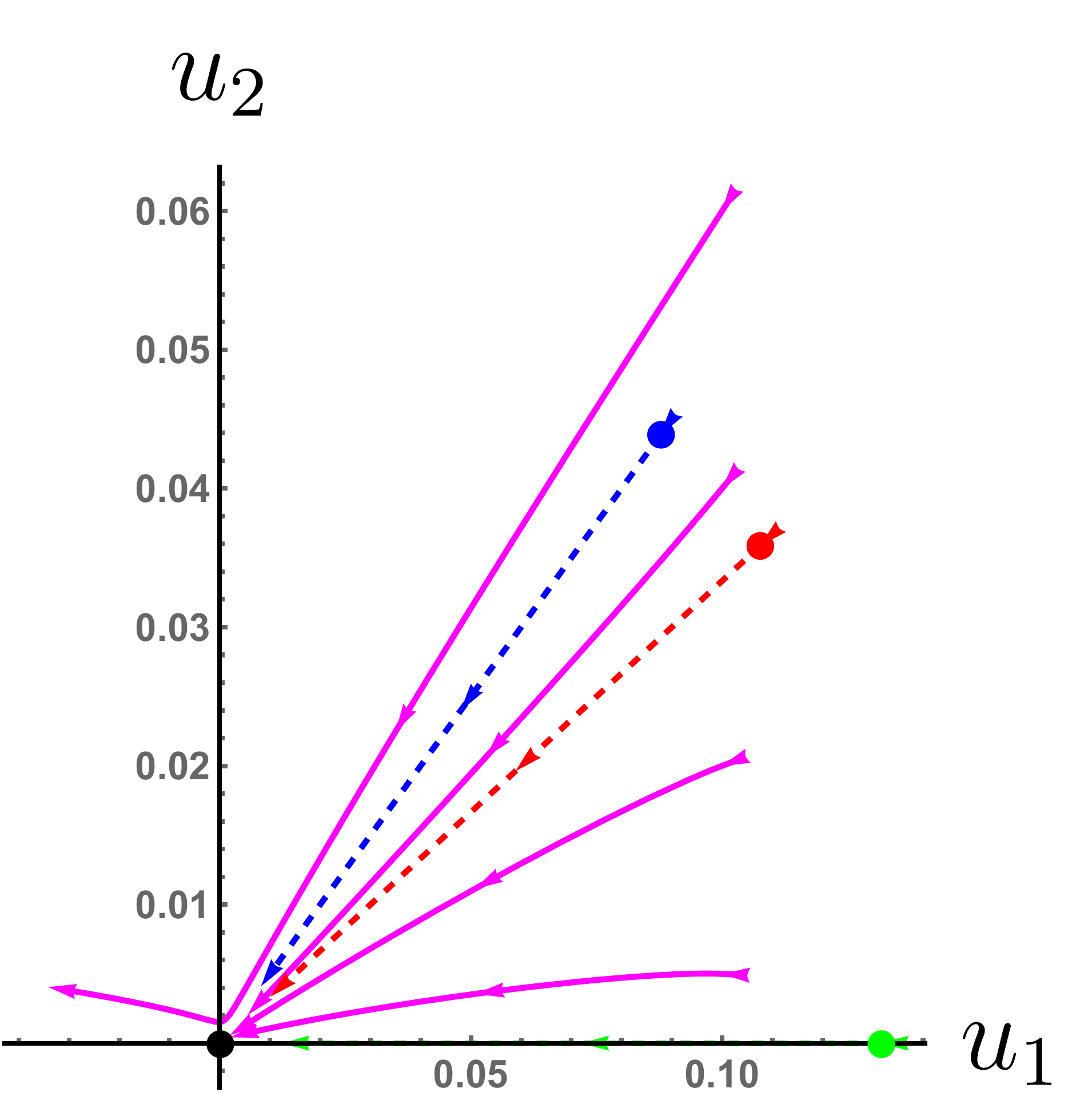}
\caption{Renormalization group flow of $u_1$ and $u_2$ when $e\neq 0$ with four different initial values of $(c,v,e,u_1,u_2)$; $(0.5, 0.3, 0.1, 0.1, 0.04)$, $(0.5, 0.1, 0.1, 0.1, 0.06)$, $(0.5, 0.1, 0.1, 0.1, 0.02)$ and $(0.5, 0.1, 0.1, 0.1, 0.005)$. Here, we also considered $N_f=1$ and $\epsilon=0.01$. The blue, red and green lines denote the flow of fixed points shown in Fig. \ref{u1_u2_flow}; $(u^*_1,u^*_2)$.}
\label{u1_u2_plot}
\end{figure}

Since $\beta_c$, $\beta_v$, and $\beta_e$ do not depend on $u_1$ and $u_2$, we analyze the renormalization group flow of $c$, $v$, and $e$ first. It is easy to figure out the renormalization group flow of the fermion velocity. $\beta_v > 0$ results in the fact that $v$ always decreases as we approach to lower energies. This confirms the emergence of perfect Fermi surface nesting in two dimensions. In other words, the fermion dynamics is localized in one direction, giving rise to effective one dimensional dynamics. According to our simple analysis, we find that $c$ also decreases as the energy scale is lowered. Not only fermions but also bosons become heavy and localized at low temperatures. Solving these three coupled renormalization group equations, we find the renormalization group flow of $c$, $v$, and $e$ as shown in Fig. \ref{c_v_e_betafunction_plot}. We point out that the coupling constant between fermions and bosons is also converging to zero as the energy scale is lowered.

Now, we consider the renormalization group flow of two kinds of boson self-interactions, $u_1$ and $u_2$, in the presence of the fermion-boson coupling, i.e., $e \neq 0$. We recall that there appear four fixed points when $e=0$, as shown in Fig. \ref{u1_u2_flow}. An essential point is that the Wilson-Fisher fixed point (green dot), the modified Wilson-Fisher fixed point (red dot), and the first order transition point (blue dot) are all proportional to $c^{2}$. As a result, they converge into the gaussian fixed point (black dot) as the boson velocity renormalizes to vanish in the low energy limit. Figure \ref{u1_u2_plot} shows the renormalization group flow of effective self-interactions, $u_1$ and $u_2$, for four different initial values in the presence of the fermion-boson interaction. The blue, red, and green dashed lines show that these effective interaction coefficients vanish to fall into the gaussian fixed point. On the other hand, when the renormalization group flow line is placed above the blue dashed line, $u_1$ and $u_2$ show the runaway renormalization group flow toward a negative value for $u_1$, which implies the fluctuation induced first-order phase transition \cite{Fluctuation_Induced_First_Order_I,Fluctuation_Induced_First_Order_II,
Fluctuation_Induced_First_Order_III,Fluctuation_Induced_First_Order_IV} as the case in the absence of the fermion-boson coupling.

The origin of this potential existence of the first order phase transition can be traced back to the nature of the fermion-boson interacting vertex. To clarify the physical mechanism of the first order phase transition, we compare the present Fermi surface problem of the CDW transition with that of the spin density wave (SDW) transition \cite{QCP_SDW}. The SDW transition may be regarded to be the O(3) symmetry version while our problem belongs to the Ising symmetry class. Comparing renormalization group equations of the present problem with those of the SDW transition \cite{QCP_SDW}, one finds that both share quite a similar structure, where most terms have their correspondences in renormalization group flow equations. However, there exists an essential different aspect between these two problems. The vertex correction for the effective interaction between electrons and order parameters gives rise to screening, reducing such interactions in the case of the $Z_2$ symmetry. On the other hand, it results in anti-screening for the case of the SDW transition. More concretely, the sign of the fermion-boson vertex function $h_3(c,v)$ [Eq. (\ref{YukawaBetaFunction})] differs from each other, where it is positive in the CDW transition while it is negative in the SDW transition. The anti-screening nature of the SDW case results in the enhancement of the fermion-boson coupling constant, which gives rise to more effective screening of boson self-interaction constants, $u_1$ and $u_2$. On the other hand, the screening nature of the CDW case reduces the screening effect for $u_1$ and $u_2$ interactions. As a result, both self-interaction parameters vanish to allow the second order phase transition in the SDW transition while the first order and the second order phase transition both seem to be able to appear in the CDW transition.
\\
\subsubsection{Beta functions for relative (dimensionless) parameters}

Even if the interaction parameter between electrons and order parameters flows to zero, this does not mean that the nature of the fixed point is Gaussian, i.e., non-interacting for itinerant electrons. In order to resolve this question, we introduce ratios of coupling parameters in the following way of $w=\frac{v}{c}$, $\lambda=\frac{e^2}{v}$, $\kappa_1=\frac{u_{1}}{c^2}$, and $\kappa_2=\frac{u_{2}}{c^2}$, respectively. The dynamical exponent $z$ and beta functions for $w$, $\lambda$, $\kappa_1$,and $\kappa_2$ are given by
\begin{gather}
z=\frac{8\pi^2 N_f}{8\pi^2N_f+\lambda w[h_2(c,cw)-h_1(c,cw)]}\\
\beta_w=\frac{\lambda w z}{16\pi^2N_f}\Big[2w[h_2(c,cw)+h_1(c,cw)]-\frac{\pi N_f}{2} \Big]\\
\beta_\lambda=z\lambda\Big[-\epsilon+\frac{\lambda}{16\pi}+\frac{w\lambda h_3(c,cw)}{8\pi^2 N_f}\Big]\\
\beta_{\kappa_1}=z\kappa_1\Big[-\epsilon+\frac{\lambda}{16\pi}+\frac{3\kappa_1}{16\pi^2}+\frac{3\kappa_2^2}{8\pi^2\kappa_1}\Big]\\
\beta_{\kappa_2}=z\kappa_2\Big[-\epsilon+\frac{\lambda}{16\pi}+\frac{\kappa_1}{8\pi^2}+\frac{5\kappa_2}{16\pi^2}\Big]
\end{gather}

Since $c$ flows to zero in the low energy limit, we consider the case of $c \rightarrow 0$. Resorting to the fact that $\lim_{c\rightarrow 0} h_1(c,cw)=\frac{\pi}{2}$, $\lim_{c\rightarrow 0} h_2(c,cw)=0$, and $\lim_{c\rightarrow 0} h_3(c,cw)=\frac{2\pi}{1+w}$, we obtain
\begin{gather}
z=\frac{8\pi^2N_f}{8\pi^2N_f-\frac{\pi \lambda w}{2}}\\
\beta_w=\frac{\lambda wz}{16\pi^2N_f}\Big[\pi w-\frac{\pi N_f}{2}\Big]\\
\beta_\lambda=z\lambda\Big[-\epsilon+\frac{\lambda}{16\pi}+\frac{\lambda}{4\pi N_f}\frac{w}{1+w}\Big]\\
\beta_{\kappa_1}=z\kappa_1\Big[-\epsilon+\frac{\lambda}{16\pi}+\frac{3\kappa_1}{16\pi^2}+\frac{3\kappa_2^2}{8\pi^2\kappa_1}\Big]\\
\beta_{\kappa_2}=z\kappa_2\Big[-\epsilon+\frac{\lambda}{16\pi}+\frac{\kappa_1}{8\pi^2}+\frac{5\kappa_2}{16\pi^2}\Big] .
\end{gather}

\begin{figure}[h]
\includegraphics[width=0.3\textwidth]{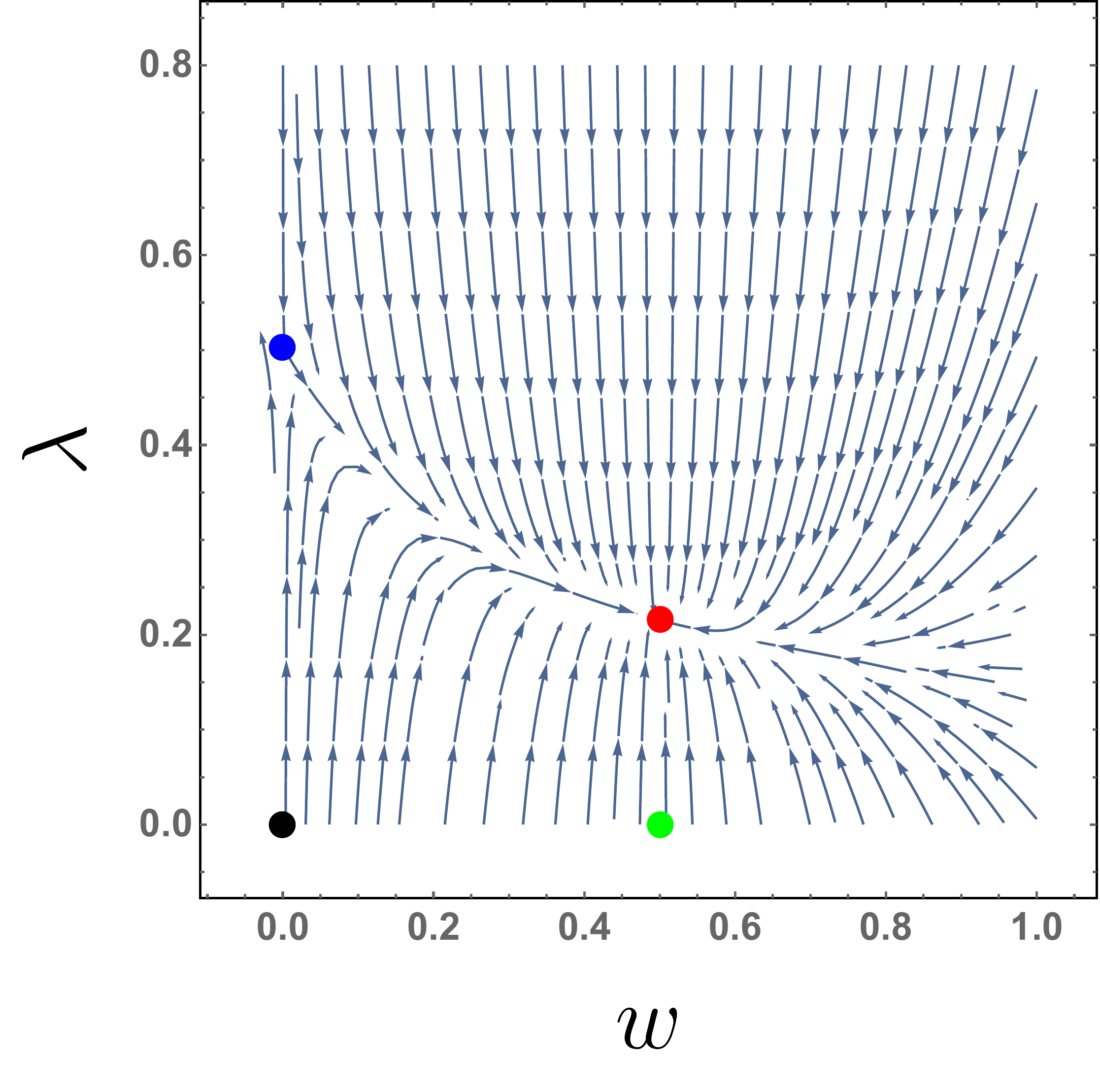}
\caption{Renormalization group flow diagram of $w$ and $\lambda$ with $N_f=1$ and $\epsilon=0.01$. Fixed points are given by the black dot $(0,0)$, the blue dot $(0,16\pi \epsilon)$, the green dot $(N_f/2,0)$, and the red dot $(N_f/2,\frac{16\pi(2+N_f)\epsilon}{6+N_f})$.}
\label{w_lambda_betafunction}
\end{figure}

\begin{figure}[h]
\includegraphics[width=0.3\textwidth]{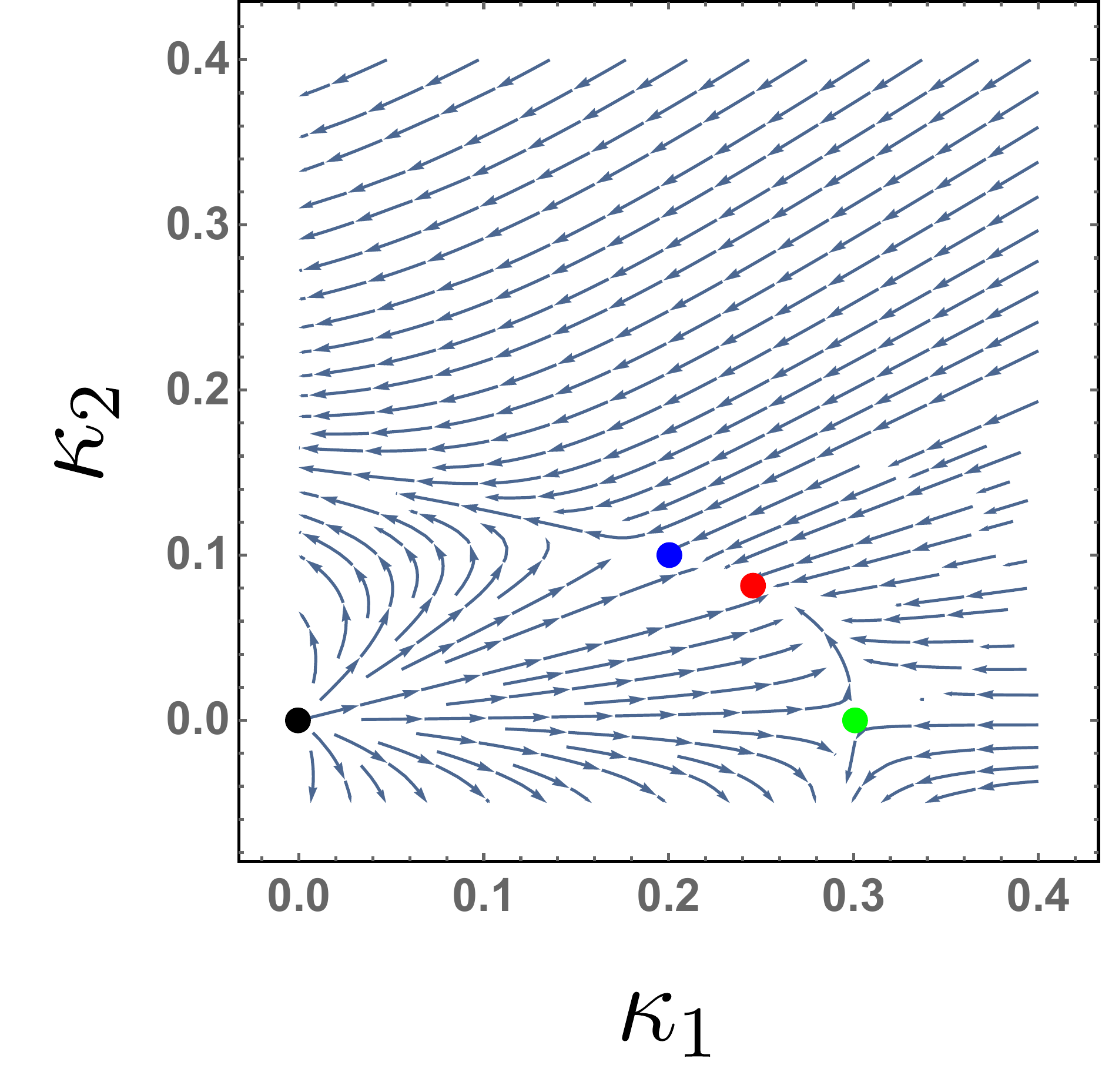}
\caption{Renormalization group flow diagram of $\kappa_{1}$ and $\kappa_{2}$ with $N_f=1$ and $\epsilon=0.01$. Fixed points are given by the black dot $(0,0)$, the green dot $(\frac{64\pi^2\epsilon}{3(6+N_f)},0)$, the red dot $(\frac{192\pi^2\epsilon}{11(6+N_f)},\frac{64\pi^2\epsilon}{11(6+N_f)})$, and the blue dot $(\frac{192\pi^2\epsilon}{9(6+N_f)},\frac{64\pi^2\epsilon}{9(6+N_f)})$.}
\label{k1_k2_betafunction}
\end{figure}

Since $z$, $\beta_w$, and $\beta_\lambda$ do not depend on $\kappa_1$ and $\kappa_2$, we analyze $\beta_w$ and $\beta_\lambda$ first. It is straightforward to find a fixed point given by $(w^*,\lambda^*) = \Big(\frac{N_f}{2},\frac{16\pi(2+N_f)\epsilon}{6+N_f}\Big)$ as shown in Fig. \ref{w_lambda_betafunction}. Putting this value to $z$, $\beta_{\kappa_1}$, and $\beta_{\kappa_2}$, we obtain
\begin{gather}
z^*=\frac{2}{2-\frac{2+N_f}{6+N_f}\epsilon}\\
\beta_{\kappa_1}=z^*\kappa_1\Big[-\frac{4}{6+N_f}\epsilon+\frac{3\kappa_1}{16\pi^2}+\frac{3\kappa_2^2}{8\pi^2\kappa_1}\Big]\\
\beta_{\kappa_2}=z^*\kappa_2\Big[-\frac{4}{6+N_f}\epsilon+\frac{\kappa_1}{8\pi^2}+\frac{5\kappa_2}{16\pi^2}\Big] .
\end{gather}
The renormalization group flow diagram for $\beta_{\kappa_1}$ and $\beta_{\kappa_2}$ is shown in Fig. \ref{k1_k2_betafunction}. There are four fixed points; one stable fixed point and three unstable fixed points. The stable fixed point (red point) is given by $(\kappa_1^*,\kappa_2^*)=(\frac{64\pi^2\epsilon}{3(6+N_f)},\frac{64\pi^2\epsilon}{11(6+N_f)})$, identified with the critical point of our CDW transition.

\section{Physical properties}

\subsection{Scaling theory of the Green's function}

Correlation functions in terms of renormalized fermion and boson fields are described by the Callan-Symanzik equation \cite{RG_Equation_Corr}, the derivation of which is shown in appendix \ref{RGdetails}. Solving Eq. (\ref{CZequation})  in appendix \ref{RGdetails}, we obtain the scaling theory for the correlation function in the vicinity of the critical point as follows
\begin{align}
&\tilde{G}_r^{(2n_f,n_b)}(\tilde{k}_0(\mu_0),\tilde{\mathbf{K}}(\mu_0),\tilde{k}_{d-1}(\mu_0),\tilde{k}_{d}(\mu_0))\nonumber\\
&=\tilde{G}_r^{2n_f,n_b}(\tilde{k}_0(\mu),\tilde{\mathbf{K}}(\mu),\tilde{k}_{d-1}(\mu),\tilde{k}_{d}(\mu))\nonumber\\
&\times\Big(\frac{\mu}{\mu_0}\Big)^{2n_f\Big(\frac{d+2}{2}-\eta_{\psi}+n_b\Big(\frac{d+3}{2}-\eta_{\Phi}+z(d-1)+2\Big)\Big)} .
\end{align}
The subscript $r$ means ``renormalized". Here, renormalized correlation functions of $2 n_{f}$ fermion fields and $n_{b}$ boson fields have been taken into account. $\tilde{k}_0(\mu)$, $\tilde{\mathbf{K}}_\perp(\mu)$, $\tilde{k}_{d-1}(\mu)$, and $\tilde{k}_d(\mu)$ are solutions of equations \ref{betafunctionsfork1} $\sim$ \ref{betafunctionsfork2} in the scaling limit, where $\mu$ is the scaling parameter. They are given by
\begin{gather}
\tilde{k}_0(\mu_0)\mu_0^{z_\tau}=\tilde{k}_0(\mu)\mu^{z_\tau} , \\
\tilde{\mathbf{K}}_\perp(\mu_0)\mu_0^{z_\perp}=\tilde{\mathbf{K}}_\perp(\mu)\mu^{z_\perp} , \\
\tilde{k}_{d-1}(\mu_0)\mu_0=\tilde{k}_{d-1}(\mu)\mu , \\
\tilde{k}_{d}(\mu_0)\mu_0=\tilde{k}_{d}(\mu)\mu
\end{gather}
near the critical point. $\eta_{\psi}$ ($\eta_{\Phi}$) is the anomalous scaling dimension of the fermion (boson) field, and $z$ is the dynamical critical exponent.

Based on this general equation for the correlation function, it is straightforward to find the scaling expression of the one-particle Green's function, given by
\begin{gather}
G_{f}(\tilde{k}_{0},\tilde{\mathbf{K}}_\perp,\tilde{k}_{d-1},\tilde{k}_{d}) = \frac{1}{|\tilde{k}_{d-1}|^{1-2\tilde{\eta}_{\Psi}}} \tilde{G}\Big(\frac{\tilde{k}_0}{|\tilde{k}_{d-1}|^{z}},\frac{\tilde{\mathbf{K}}_\perp}{|\tilde{k}_{d-1}|^z}\Big)
\end{gather}
for the fermion propagator of the hot spot $+1$ and
\begin{gather}
G_{b}(\tilde{k}_{0},\tilde{\mathbf{K}}_\perp,\tilde{k}_{d-1},\tilde{k}_{d}) = \frac{C}{(\tilde{k}_0^2+|\tilde{\mathbf{K}}_\perp|^2)^{\frac{2-2\tilde{\eta}_\Phi}{4}}}
\end{gather}
for the boson Green's function. Here, C is a positive constant. We have anomalous scaling for both fermion and boson Green's functions, characterized by $\tilde{\eta}_{\Psi}=\eta_{\Psi}+\frac{(2-\epsilon)(z-1)}{2}$ and
%
%
$\tilde{\eta}_\Phi=\eta_\Phi+\frac{(2-\epsilon)(z-1)}{2}$, where
%
%
both fermion and boson anomalous scaling dimensions are given by
\begin{gather}
%
%
\eta_\Psi=\eta_{\Phi}=-\frac{1}{2}\frac{2+N_f}{6+N_f}\epsilon
\end{gather}
up to the $\mathcal{O}(\epsilon)$ order. The dynamical critical exponent is
\begin{gather}
z=1+\frac{1}{2}\frac{2+N_f}{6+N_f}\epsilon .
%
%
\end{gather}
We would like to emphasize that the fermion Green's function does not depend on $k_{d}$ and the boson propagator does not rely on $k_{d-1}$ and $k_d$ in the low energy limit. This scaling theory originates from the fact that the fermion velocity $v$ and the boson velocity $c$ go to vanish in the low energy limit.

\subsection{Enhancement of Fermi surface nesting in two dimensions}

\begin{figure}[h]
\includegraphics[width=0.3\textwidth]{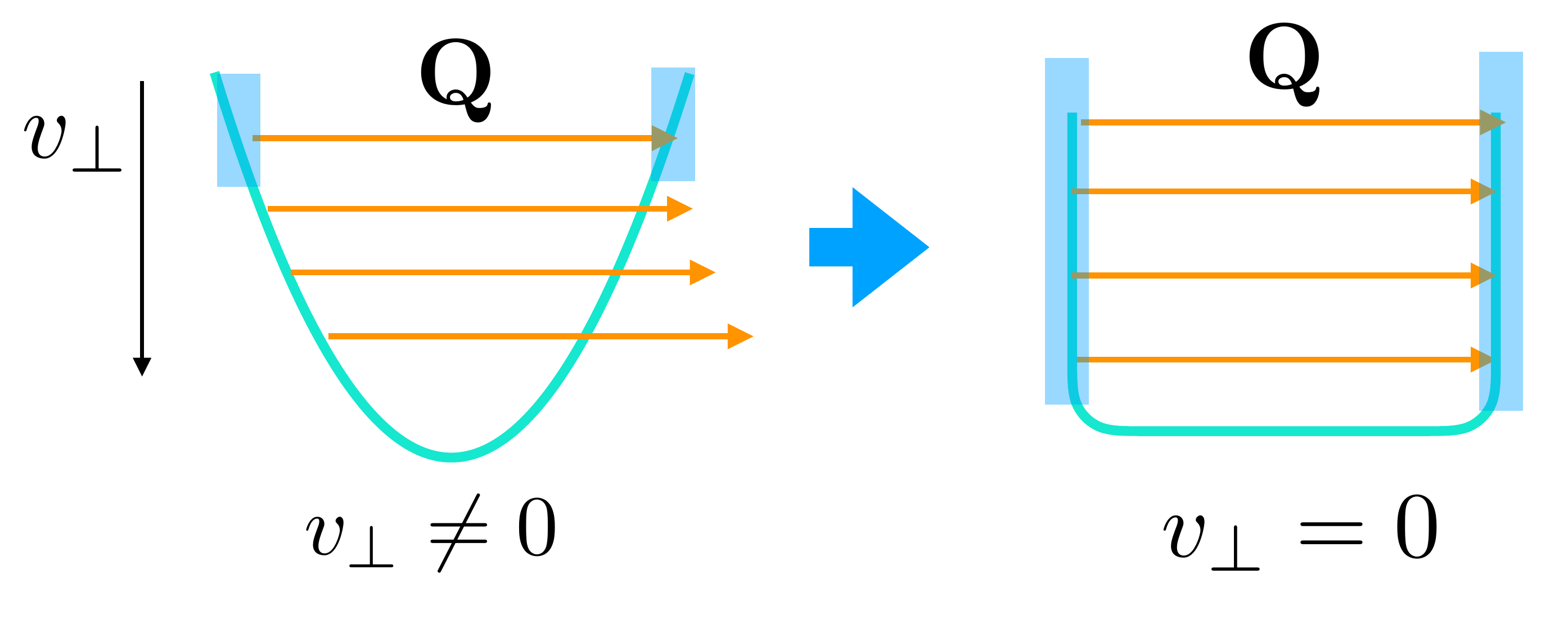}
\caption{Schematic description on how Fermi surface nesting is enhanced by electron-electron correlations. Here, the blue regime refers to the phase space connected by the nesting vector $\mathbf{Q}$ (orange arrow). $v_{\perp}$ is one component of the Fermi velocity, orthogonal to the nesting vector. As $v_{\perp}$ approaches to the zero value, the Fermi surface nesting becomes much stronger.}
\label{PerfectNesting}
\end{figure}

Our beta function analysis showed that the fermion velocity perpendicular to the nesting vector $\mathbf{Q}_i$ decreases as we approach to lower energies. It means that there appear effective Fermi lines which can be more connected by the nesting vector $\mathbf{Q}_i$ at low energies, as shown in Fig. \ref{PerfectNesting}. However, we emphasize that this occurs away from three dimensions. More precisely, we find the fermion velocity as a function of an energy scale $\mu$ with the dimensional regularization parameter $\epsilon$
\begin{align*}
&\epsilon\neq 0:\;\; v(\mu)\sim\frac{v_0}{\sqrt{1-\ln\Big(\frac{\mu}{\mu_0}\Big)}} \\
&\epsilon= 0:\;\; v(\mu)\sim \frac{v_0}{\sqrt{1+\ln\Big[\ln\Big(\frac{\mu_0}{\mu}\Big)\Big]}}.
\end{align*}
This different form of $v(\mu)$ originates from whether $\lambda = \frac{e^2}{v}$ flows to zero or non-zero, respectively.

\begin{figure}[h]
\includegraphics[width=0.35\textwidth]{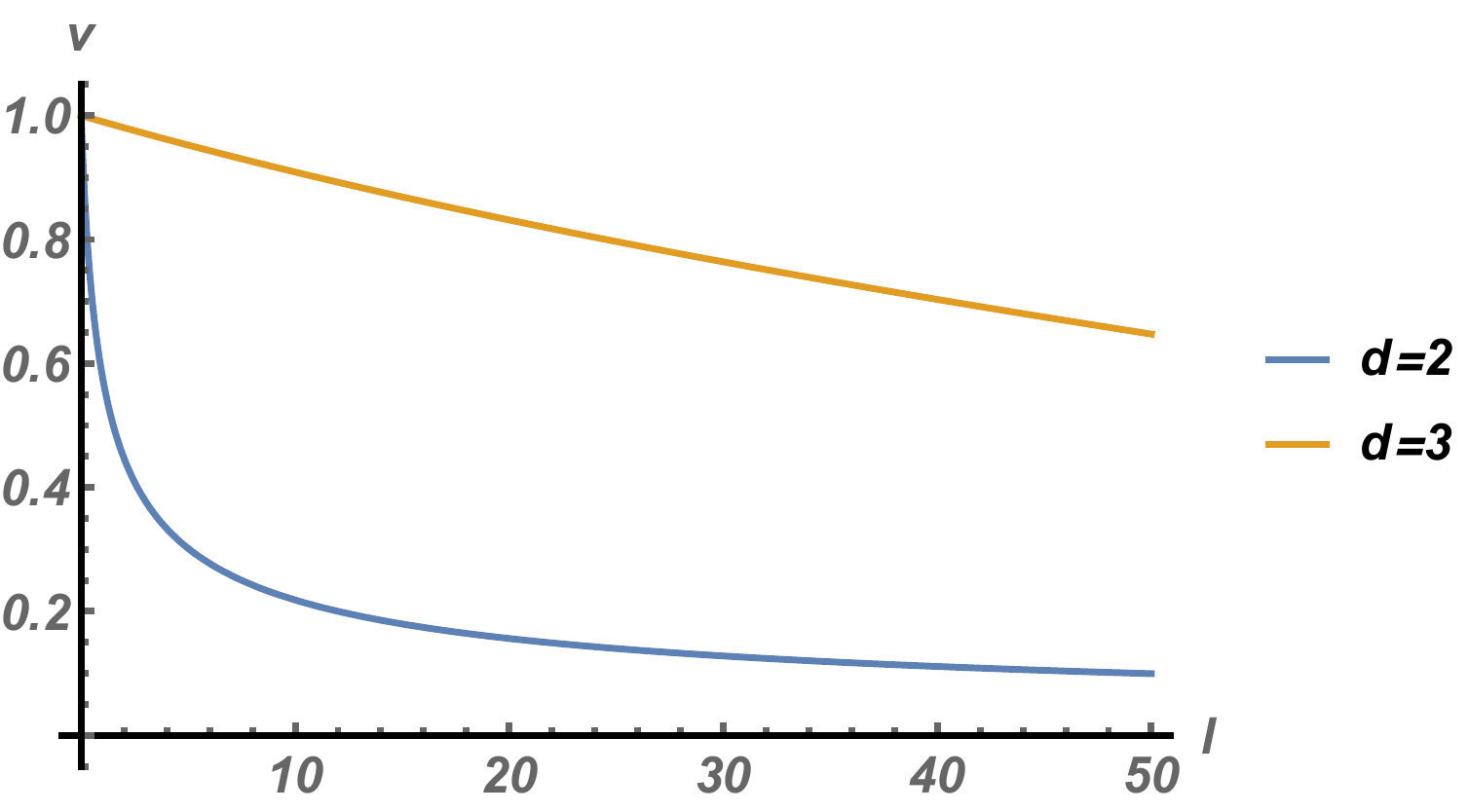}
\caption{Evolution of the fermion velocity $v(l)$ as approaching to lower energies ($l\uparrow$) for two different cases; $\epsilon=0$ $(d=3)$ and $\epsilon=1$ $(d=2)$.} \label{vflow_2dvs3d}
\end{figure}

Fig. \ref{vflow_2dvs3d} shows $v(l)$ as $l(=\ln(\mu_0/\mu))$ increases, i.e., energy decreased, for two cases; $d=3$ $(\epsilon=0)$ and $d=2$ $(\epsilon=1)$. $v$ decreases faster in $d=2$, compared to the case of $d=3$. It means that the two dimensional system is more favorable for the nesting effect than three dimensional systems. However, this argument is not correct completely. Since we used the co-dimensional regularization, the Fermi surface remains to be one dimensional line even in the three dimensional case. It is different from the real Fermi surface of this system. The actual Fermi surface for the $VSe_2$ bulk system has been well known \cite{CDW_VSe2_Bulk_I,CDW_VSe2_Bulk_II,CDW_VSe2_Bulk_III}. Although we cannot give quantitative analysis for this real case, we can deduce some rigorous statements based on the scaling analysis. Since three dimensional $VSe_2$ is basically given by stacking of two dimensional $VSe_2$ layers, it is weakly dispersive along the stacking direction. This can be confirmed in the actual Fermi surface which turns out to be rather ``flat" along the stacking direction. Based on this discussion, we assume that the dispersion relation for hot electrons is given by $\epsilon(k_1,k_2,k_3)\sim k_1+vk_2+\sum_{n>1}k_3^n$, where $k_3$ is the coordinate along the stacking direction. As a result, we deduce the scaling dimension of the coupling constant $e$: $[e]=\frac{1}{2}\Big(\frac{1}{n}-1\Big)$. This scaling relation gives rise to $[e]<0$ as long as $n>1$, which implies that $e$ is always irrelevant in the low energy regime. Since $e$ is essential in the renormalization of $v$ as shown in $\beta_v$, we conclude that the perfect Fermi-surface nesting would not occur in three dimensional $VSe_{2}$. This argument is consistent with why the Hertz-Moriya-Millis theory \cite{HMM_EFT_QCP_I,HMM_EFT_QCP_II,HMM_EFT_QCP_III,HMM_EFT_QCP_IV} works well in three dimensions although it breaks down in two dimensions. 
%
%

One may criticize that the emergence of perfect nesting of the Fermi surface results from just reducing the dimensionality and involved with the lattice structure. Actually, this is related with the main point of the present study, reflected in the title. Although the initial Fermi surface nesting structure is not ``good" at high temperatures (UV), there exists a temperature evolution for the Fermi velocity near CDW criticality in two dimensions: The Fermi surface nesting becomes perfect at low temperatures (IR), entitled with ``Universal renormalization group flow toward perfect Fermi-surface nesting near CDW criticality in two dimensions." On the other hand, this renormalization group flow, i.e., the temperature evolution of the Fermi surface structure does not occur near criticality in three dimensions as discussed above. Furthermore, if the ``approximately" perfect Fermi surface nesting is regarded to be just an effect of the band structure involved with the dimensional reduction, there do not exist temperature evolutions toward perfect Fermi surface nesting.

Suppose that the Fermi surface nesting property is quite nice in either three or two dimensions, described by a band structure calculation. This can happen either accidentally or inevitably, which can originate from interference effects due to the lattice structure. An important point is that there is no such strong renormalization group flow, i.e., temperature evolution for the Fermi surface structure both near CDW criticality in three dimensions and in the band structure effect of two dimensions, where the band structure does not change much from high temperatures to low temperatures. However, the Fermi velocity acquires strong renormalization effects from enhanced interactions between electrons, responsible for the emergence of the perfect Fermi surface nesting in two dimensions. This temperature evolution may be resolved in the ARPES experiment, which requires high-precision energy and momentum resolution. Unfortunately, recent experiments did not verify this issue \cite{ARPES_STM_CDW_MIT}.

\section{Beyond the one-loop order: Discussion on controllability}

To check the validity of our renormalization group analysis up to the one-loop order, we investigate the controllability of the present Fermi-surface problem, following the S.-S. Lee's paper \cite{QCP_SDW}. For a generic Feynman diagram, the amplitude of the diagram $I$ is given by
\begin{align*}
I&\sim e^{V_e}u_i^{V_u}\int\Big[\prod_{i=1}^{L}dp_i\Big]\prod_{j=1}^{I_f}\Big\{\frac{1}{\mathbf{\Gamma}\cdot\mathbf{K}_j+\gamma_{d-1}[\pm k_{d-1,j}+vk_{d,j}]}\Big\}\\
&\times\prod_{l=1}^{I_b}\Big[\frac{1}{|\mathbf{Q}_l|^2+c^2(q_{d-1,l}^2+q_{d,l}^2)}\Big] .
\end{align*}
Here, $V_{e}$ and $V_{u}$ are the number of interaction vertices for $e$ and $u$, respectively. $I_{f}$ ($I_{b}$) is the number of fermion (boson) propagators, and $L$ is the total number of loops. $k$ and $q$ are momentum of fermions and bosons, respectively, which consist of loop momentum $p$ and external momentum. If we denote the external momentum as $P_{ext}$, we obtain $\{k\}=\{\alpha_k\}p+\{\beta_k\}P_{ext}$ and $\{q\}=\{\alpha_q\}p+\{\beta_q\}P_{ext}$.

First, we transform $p_d$ into $\frac{1}{v} p_d$. Under this transformation, we obtain
\begin{gather}
k_{d,j}\rightarrow \frac{1}{v}(\alpha_{k_d,j}p_d+v\beta_{k_d,j}P_{ext,d})\equiv \frac{1}{v}k'_{d,j}\\
q_{d,j}\rightarrow \frac{1}{v}(\alpha_{q_d,j}p_d+v\beta_{q_d,j}P_{ext,d})\equiv \frac{1}{v}q'_{d,j} .
\end{gather}
Then, we rewrite the above expression as follows
\begin{align*}
I&\sim \frac{e^{V_e}u_i^{V_u}}{v^{L}}\int\Big[\prod_{i=1}^{L}dp_i\Big]\prod_{j=1}^{I_f}\Big\{\frac{1}{\mathbf{\Gamma}\cdot\mathbf{K}_j+\gamma_{d-1}[\pm k_{d-1,j}+k'_{d,j}]}\Big\}\\
&\times\prod_{l=1}^{I_b}\Big[\frac{1}{|\mathbf{Q}_l|^2+c^2q_{d-1,l}^2+\frac{1}{w^2}q'^2_{d,l}}\Big]
\end{align*}
The above integral converges when there are no loops which consist of only bosonic propagators. If there are loops which consist of only bosonic propagators, this gives rise to the divergence in the zero $c$ limit. In other words, this loop integral is proportional to $\frac{1}{c^{L_b}}$, where $L_{b}$ is the number of loops consisting of only boson propagators. As a result, we reach the following expression
\begin{align}
I\sim \frac{e^{V_e}u_i^{V_u}}{v^Lc^{L_b}}=w^{-V_u}\lambda^{\frac{V_e+2-E}{2}}\kappa_i^{V_u}e^{E-2}c^{-L_b+V_u} ,
\end{align}
where $E$ is the number of external lines. In the above, $L+V-I=1$ with $V = V_e + V_u$ and $I = I_f + I_b$, and $3V_e+4V_u=2I+E$ have been utilized. Based on this result, we obtain the magnitude of renormalization constants for the propagator $(E=2)$, the Yukawa coupling vertex $(E=3)$, and the quartic vertex $(E=4)$
\begin{align*}
I_{E=2}&\sim w^{-V_u}\lambda^{\frac{V_e}{2}}\kappa_i^{V_u}c^{\delta} , \\
I_{E=3}&\sim w^{-V_u}\lambda^{\frac{V_e-1}{2}}\kappa_i^{V_u}c^{\delta}e , \\
I_{E=4}&\sim w^{-V_u}\lambda^{\frac{V_e-2}{2}}\kappa_i^{V_u}e^2c^{\delta} ,
\end{align*}
where $\delta=V_u-L_b \geq 0$.

To find the $A_i$ coefficients in counterterms, we should also consider some coefficients in front of $A_i$. Then, we obtain
\begin{align*}
A_0,A_1&\sim \frac{\partial I_{E=2}}{\partial P_{ext,0}}\sim \frac{\partial I_{E=2}}{\partial P_{ext,\perp}}\sim w^{-V_u}\lambda^{\frac{V_e}{2}}\kappa_i^{V_u}c^{\delta} , \\
A_2&=\frac{\partial I_{E=2}}{\partial P_{ext,{d-1}}}\sim w^{-V_u}\lambda^{\frac{V_e}{2}}\kappa_i^{V_u}c^{\delta}(1+c^2) , \\
A_3&=\frac{1}{v}\frac{\partial I_{E=2}}{\partial P_{ext,{d}}}\sim w^{-V_u}\lambda^{\frac{V_e}{2}}\kappa_i^{V_u}c^{\delta}(1+w^{-2}) , \\
A_4,A_5&\sim \frac{\partial^2 I_{E=2}}{\partial P_{ext,{0}}^2}\sim \frac{\partial^2 I_{E=2}}{\partial P_{ext,{\perp}}^2}\sim w^{-V_u}\lambda^{\frac{V_e}{2}}\kappa_i^{V_u}c^{\delta} , \\
A_6&\sim \frac{1}{c^2}\Big(\frac{\partial^2 I_{E=2}}{\partial P_{ext,{d-1}}^2}+\frac{\partial^2 I_{E=2}}{\partial P_{ext,{d}}^2}\Big) , \\
&\sim  w^{-V_u}\lambda^{\frac{V_e}{2}}\kappa_i^{V_u}c^{\delta}(c^{-2}+1+c^2+w^2+w^{-2}) , \\
A_7&\sim \frac{1}{e}I_{E=3}\sim w^{-V_u}\lambda^{\frac{V_e-1}{2}}\kappa_i^{V_u}c^{\delta} , \\
A_8,A_9&\sim \frac{1}{u_i}I_{E=4} \sim w^{-V_u+1}\lambda^{\frac{V_e}{2}}\kappa_i^{V_u-1}c^{\delta-1} .
\end{align*}
Up to the one-loop order, we find $w\sim\mathcal{O}(1)$ and $\lambda, ~\kappa_i \sim \mathcal{O}(\epsilon)$, where $c$ goes to the zero limit. Therefore, higher-order loop contributions to $A_0,\cdots,A_5$ and $A_7$ can be ignored in the small $\epsilon$ limit. On the other hand, such higher-order loop contributions to $A_6$ and $A_8$, $A_9$ cannot be neglected when $\delta$ is smaller than two for $A_6$ and one for $A_8$, $A_9$, respectively, in the limit of non-zero $\epsilon$ since $c$ goes to zero in the low energy limit. Fortunately, even these terms can be ignored in three dimensions since $c$ converges to zero slower than $\lambda$ and $\kappa$. We reall $c\sim \frac{1}{(\ln l)^{1/2}}$, $\lambda, ~\kappa_i\sim\frac{1}{l}$, where $l\sim\ln\mu$. However, such higher-order diagrams should be taken into account for two dimensional systems.
	
\section{Summary}

Dimensionality and hetero-interface structure of quantum material are essential factors to control both electron-electron and electron-phonon interactions, responsible for electronic reconstruction phenomena, which serves as the basic principle for device applications. Compared with the electronic reconstruction paradigm in oxide hetero-structured quantum materials, such phenomena appear as rather a simple fashion in the van der Waals hetero-interface system, thus expected to be an ideal flat form testing the basic principle in the strongly correlated regime, for example, metallic quantum criticality in two dimensions. Actually, recent ARPES and STM measurements demonstrated that physics of strong correlations arises in monolayer $VSe_{2}$ \cite{ARPES_STM_CDW_MIT}. In particular, the ARPES experiment has shown perfect Fermi-surface nesting, implying further dimensional reduction that one dimensional motion of electrons is realized instead of two dimensional dynamics.

In order to understand this strongly correlated dynamics of electrons, we constructed an effective field theory in terms of itinerant electrons and CDW critical fluctuations. Resorting to a novel dimensional regularization technique for this Fermi surface problem \cite{QCP_Dimensional_Regularization,Mott_Dimensional_Regularization,QCP_SDW}, we performed the renormalization group analysis to reveal the mechanism for perfect Fermi surface nesting in the monolayer $VSe_{2}$ system. The renormalization group flow for the curvature parameter gives rise to the emergence of the perfect Fermi surface nesting universally only in two dimensions beyond the Hertz-Moriya-Millis description in three dimensions \cite{HMM_EFT_QCP_I,HMM_EFT_QCP_II,HMM_EFT_QCP_III,HMM_EFT_QCP_IV}. We claim that this further dimensional reduction from the two dimensional Fermi surface with imperfect Fermi surface nesting to the one dimensional Fermi surface with perfect Fermi surface nesting is responsible for the drastic enhancement of the CDW ordering transition temperature although the CDW ordering itself follows that of the bulk parent. We point out that this further dimensional reduction in the dynamics of electrons has been also reported in two dimensional SDW transitions \cite{QCP_SDW}.

\section*{ACKNOWLEDGEMENT}

This study was supported by the Ministry of Education, Science, and Technology (No. NRF-2015R1C1A1A01051629 and No. 2011-0030046) of the National Research Foundation of Korea (NRF). I.J. is supported by Global Ph.D. Fellowship of the National Research Foundation of Korea(NRF-2015H1A2A1033126).

\widetext

\appendix

\section{Preparation for renormalization group analysis} \label{RGdetails}

\subsection{Counterterms and renormalized effective field theory}

We start from an effective bare action given by
\begin{align}
S_{b,0}&=\sum_{n=1}^3\sum_{m=\pm}\sum_{j=1}^{N_f}\int dk_b\bar{\Psi}_{b,n,j}^{(m)}(k_b)[i\gamma_0k_{b,0}+i\mathbf{\Gamma}_\perp\cdot\mathbf{K}_{b,\perp}+i\gamma_{d-1}\epsilon_n^{(m)}(k_{b,d-1},k_{b,d},v_b)]\Psi_{b,n,j}^{(m)}(k)\nonumber \\
&+\frac{1}{2}\sum_{n=1}^3\int dk_b [|k_0|^2+|\mathbf{K}_{b,\perp}|^2+c_b^2(k_{b,d-1}^2+k_{b,d}^2)]\Phi_{b,n}(k_b)\Phi_{b,n}(-k_b)\\
S_{b,int-bf}&=\frac{ie_b}{\sqrt{N_f}}\sum_{n=1}^{3}\sum_{j=1}^{N_f}\int dk_b\int dq_b \Phi_{b,n}(q_b)\Big[\bar{\Psi}_{b,n,j}^{(-)}(k_b+q_b)\gamma_{d-1}\Psi_{b,n,j}^{(+)}(k_b)+\bar{\Psi}_{b,n,j}^{(+)}(k_b+q_b)\gamma_{d-1}\Psi_{b,n,j}^{(-)}(k_b)\Big)\Big]\\
S_{b,int-b1}&=\frac{u_{1b}}{4!}\sum_{n=1}^3\int \prod_{i=1}^4dq_b\Phi_{b,n}(q_{b,1})\Phi_{b,n}(q_{b,2})\Phi_{b,n}(q_{b,3})\Phi_{b,n}(q_{b,4})(2\pi)^{d+1}\delta(q_{b,1}+q_{b,2}+q_{b,3}+q_{b,4})\\
S_{b,int-b2}&=\frac{u_{2b}}{2!2!}\int\prod_{i=1}^4dq_{b,i}\Big[\Phi_{b,1}(q_{b,1})\Phi_{b,1}(q_{b,2})\Phi_{b,2}(q_{b,3})\Phi_{b,2}(q_{b,4})+\Phi_{b,2}(q_{b,1})\Phi_{b,2}(q_{b,2})\Phi_{b,3}(q_{b,3})\Phi_{b,3}(q_{b,4})\nonumber\\
&+\Phi_{b,3}(q_{b,1})\Phi_{b,3}(q_{b,2})\Phi_{b,1}(q_{b,3})\Phi_{b,1}(q_{b,4})\Big](2\pi)^{d+1}\delta(q_{b,1}+q_{b,2}+q_{b,3}+q_{b,4}) .
\end{align}

Introducing quantum corrections into this effective field theory, various ultraviolet (UV) divergences appear. Such UV divergences are canceled by so called counterterms

\begin{align}
S_{ct,0}&=\sum_{n=1}^3\sum_{m=\pm}\sum_{j=1}^{N_f}\int dk_b\bar{\Psi}_{r,n,j}^{(m)}(k_r)[iA_0\gamma_0k_{r,0}+iA_1\mathbf{\Gamma}_\perp\cdot\mathbf{K}_{r,\perp}+iA_2\gamma_{d-1}\epsilon_n^{(m)}(k_{r,d-1},k_{r,d},\frac{A_3}{A_2}v_r)]\nonumber \\
&\times\Psi_{r,n,j}^{(m)}(k)+\frac{1}{2}\sum_{n=1}^3\int dk_r [A_4|k_0|^2+A_5|\mathbf{K}_{r,\perp}|^2+A_6c_r^2(k_{r,d-1}^2+k_{r,d}^2)]\Phi_{r,n}(k_r)\Phi_{r,n}(-k_r)\\
S_{ct,int-bf}&=\frac{iA_7\tilde{e}_r\mu^{\epsilon/2}}{\sqrt{N_f}}\sum_{n=1}^{3}\sum_{j=1}^{N_f}\int dk_r\int dq_r \Phi_{r,n}(q_r)\Big[\bar{\Psi}_{r,n,j}^{(-)}(k_r+q_r)\gamma_{d-1}\Psi_{r,n,j}^{(+)}(k_r)
+\bar{\Psi}_{r,n,j}^{(+)}(k_r+q_r)\gamma_{d-1}\Psi_{r,n,j}^{(-)}(k_r)\Big)\Big]\\
S_{ct,int-b1}&=\frac{A_8\tilde{u}_{1r}\mu^{\epsilon}}{4!}\sum_{n=1}^3\int \prod_{i=1}^4dq_r\Phi_{r,n}(q_{r,1})\Phi_{r,n}(q_{r,2})\Phi_{r,n}(q_{r,3})\Phi_{r,n}(q_{r,4})(2\pi)^{d+1}\delta(q_{r,1}+q_{r,2}+q_{r,3}+q_{r,4})\\
S_{ct,int-b2}&=A_9\frac{\tilde{u}_{2r}\mu^{\epsilon}}{2!2!}\int\prod_{i=1}^4dq_{r,i}\Big[\Phi_{r,1}(q_{r,1})\Phi_{r,1}(q_{r,2})\Phi_{r,2}(q_{r,3})\Phi_{r,2}(q_{r,4})+\Phi_{r,2}(q_{r,1})\Phi_{r,2}(q_{r,2})\Phi_{r,3}(q_{r,3})\Phi_{r,3}(q_{r,4})\nonumber\\
&+\Phi_{r,3}(q_{r,1})\Phi_{r,3}(q_{r,2})\Phi_{r,1}(q_{r,3})\Phi_{r,1}(q_{r,4})\Big](2\pi)^{d+1}\delta(q_{r,1}+q_{r,2}+q_{r,3}+q_{r,4}) ,
\end{align}
where UV divergences are absorbed into $A_{n}$ coefficients with $n = 0, ..., 9$.

Extracting all counterterms from the bare action, i.e., $S_{r} = S_{b} - S_{ct}$, we have an effective renormalized action, given by

\begin{align}
S_{r,0}&=\sum_{n=1}^3\sum_{m=\pm}\sum_{j=1}^{N_f}\int dk_b\bar{\Psi}_{r,n,j}^{(m)}(k_r)[i\gamma_0k_{r,0}+i\mathbf{\Gamma}_\perp\cdot\mathbf{K}_{r,\perp}+i\gamma_{d-1}\epsilon_n^{(m)}(k_{r,d-1},k_{r,d},v_r)]\Psi_{r,n,j}^{(m)}(k)\nonumber \\
&+\frac{1}{2}\sum_{n=1}^3\int dk_r [|k_0|^2+|\mathbf{K}_{r,\perp}|^2+c_r^2(k_{r,d-1}^2+k_{r,d}^2)]\Phi_{r,n}(k_r)\Phi_{r,n}(-k_r)\\
S_{r,int-bf}&=\frac{i\tilde{e}_r\mu^{\epsilon/2}}{\sqrt{N_f}}\sum_{n=1}^{3}\sum_{j=1}^{N_f}\int dk_r\int dq_r \Phi_{r,n}(q_r)\Big[\bar{\Psi}_{r,n,j}^{(-)}(k_r+q_r)\gamma_{d-1}\Psi_{r,n,j}^{(+)}(k_r)+\bar{\Psi}_{r,n,j}^{(+)}(k_r+q_r)\gamma_{d-1}\Psi_{r,n,j}^{(-)}(k_r)\Big)\Big]\\
S_{r,int-b1}&=\frac{\tilde{u}_{1r}\mu^{\epsilon}}{4!}\sum_{n=1}^3\int \prod_{i=1}^4dq_r\Phi_{r,n}(q_{r,1})\Phi_{r,n}(q_{r,2})\Phi_{r,n}(q_{r,3})\Phi_{r,n}(q_{r,4})(2\pi)^{d+1}\delta(q_{r,1}+q_{r,2}+q_{r,3}+q_{r,4})\\
S_{r,int-b2}&=\frac{\tilde{u}_{2r}\mu^{\epsilon}}{2!2!}\int\prod_{i=1}^4dq_{r,i}\Big[\Phi_{r,1}(q_{r,1})\Phi_{r,1}(q_{r,2})\Phi_{r,2}(q_{r,3})\Phi_{r,2}(q_{r,4})+\Phi_{r,2}(q_{r,1})\Phi_{r,2}(q_{r,2})\Phi_{r,3}(q_{r,3})\Phi_{r,3}(q_{r,4})\nonumber\\
&+\Phi_{r,3}(q_{r,1})\Phi_{r,3}(q_{r,2})\Phi_{r,1}(q_{r,3})\Phi_{r,1}(q_{r,4})\Big](2\pi)^{d+1}\delta(q_{r,1}+q_{r,2}+q_{r,3}+q_{r,4}) ,
\end{align}
where UV divergences disappear and well defined. Here, we introduce an energy scale $\mu$ to make $e_r$, $u_{1r}$, and $u_{2r}$ be dimensionless quantities, redefined by $\tilde{e}_r$, $\tilde{u}_{1r}$, and $\tilde{u}_{2r}$. The upper critical dimension for all interaction parameters of $\tilde{e}_r$, $\tilde{u}_{1r}$, and $\tilde{u}_{2r}$ turns out to be $d_{c} = 3$, where the expansion parameter is given by $\epsilon = 3 - d$ in the dimensional regularization scheme.

It is straightforward to find equations between bare and renormalized quantities. First, we consider the scaling transformation, given by
\begin{gather}
k_{b,0}=Z_\tau k_{r,0},\;\;\; \mathbf{K}_{b,\perp}=Z_\perp \mathbf{K}_{r,\perp}, \nonumber \\ k_{b,d-1}=k_{r,d-1},\;\;\; k_{b,d}=k_{r,d} . \label{bare-renor2}
\end{gather}
Here, $Z_\tau$ and $Z_\perp$ are rescaling parameters for frequency and ``transverse" momentum in fictitious extra dimensions. Second, we introduce field renormalization constants of $Z_\Psi$ and $Z_\Phi$, which relate bare fields with renormalized ones in the following way
\begin{gather}
\Psi_b=Z_\Psi^{1/2}\Psi_r,\;\;\; \Phi_b=Z_\Phi^{1/2}\Phi_r . \label{bare-renor1}
\end{gather}
Then, resorting to $S_{b}=S_{r}+S_{ct}$, we define all renormalized constants $Z_{n}$ with $n = 0, ..., 9$
\begin{gather}
Z_\Psi Z_{\perp}^{d-2}Z_\tau^2=Z_0,\; Z_\Psi Z_\perp^{d-1}Z_\tau=Z_1, \nonumber \\ Z_\Psi Z_\perp^{d-2}Z_\tau=Z_2,\; Z_\Psi Z_\perp^{d-2}Z_\tau v_b=Z_3v_r\\
Z_\Phi Z_\perp^{d-2}Z_\tau^3=Z_4,\; Z_\Phi Z_\perp^{d}Z_\tau =Z_5, \nonumber \\ Z_\Phi Z_{\perp}^{d-2}Z_\tau c_b^2=Z_6 c_r^2, \\ Z_\Phi^{1/2}Z_\Psi Z_\perp^{2(d-2)}Z_\tau^2e_b=Z_7 \tilde{e}_r\mu^{\epsilon/2}\\
Z_\Phi^2Z_{\perp}^{3(d-2)}Z_\tau^3u_{1b}=Z_8\tilde{u}_{1r}\mu^\epsilon, \nonumber \\ Z_\Phi^2Z_{\perp}^{3(d-2)}Z_\tau^3u_{2b}=Z_9\tilde{u}_{2r}\mu^\epsilon ,
\end{gather}
where such renormalization constants are given by counterterm coefficients as
\begin{gather}
Z_n=1+A_n .
\end{gather}

\subsection{Feynman rules}

In order to perform the perturbative renormalization group analysis systematically, we introduce Feynman rules based on the renormalized effective action and counterterms. Here, we express the fermion-involved sector in a more compact way as follows:

\begin{align}
S_{r,0}&=\sum_{n=1}^3\sum_{j=1}^{N_f}\int dk_r \bar{\Psi}_{r,n,j}(k_r)\left(\begin{array}{cc} i\mathbf{\Gamma}\cdot\mathbf{K}_r+i\gamma_{d-1}\epsilon_n^{(-)} & 0 \\ 0 &i\mathbf{\Gamma}\cdot\mathbf{K}_r+i\gamma_{d-1}\epsilon_n^{(+)} \end{array}\right)\Psi_{r,n,j}(k_r)\nonumber \\
&+\frac{1}{2}\sum_{n=1}^3\int dk_r[|\mathbf{K}_r|^2+c_r^2(k_{r,d-1}^2+k_{r,d}^2)]\Phi_{r,n}(k_r)\Phi_{r,n}(-k_r)\\
S_{r,int-bf}&=i\frac{\tilde{e}_r\mu^{\epsilon/2}}{\sqrt{N_f}}\sum_{n=1}^3\sum_{j=1}^{N_f}\int dk_r\int dq_r\bar{\Psi}_{r,n,j}(k_r+q_r)\Phi_{r,n}(q_r)\gamma_{d-1}\otimes \sigma_1\Psi_{r,n,j}(k_r)\\
S_{ct,0}&=\sum_{n=1}^3\sum_{j=1}^{N_f}\int dk_r\bar{\Psi}_{r,n,j}\nonumber \\
&\left(\begin{array}{cc}iA_0k_{r,0}+iA_1\mathbf{\Gamma}_\perp\cdot\mathbf{K}_{r,\perp}+iA_2\gamma_{d-1}\epsilon_n^{(-)}(\frac{A_3}{A_2}v_r) & 0 \\ 0 & iA_0k_{r,0}+iA_1\mathbf{\Gamma}_\perp\cdot\mathbf{K}_{r,\perp}+iA_2\gamma_{d-1}\epsilon_n^{(+)}(\frac{A_3}{A_2}v_r)\end{array}\right)\nonumber \\
&\times \Psi_{r,n,j}+\frac{1}{2}\sum_{n=1}^3\int dk_r[A_4|k_{r,0}|^2+A_5|\mathbf{K}_{r,\perp}|^2+A_6c_r^2(k_{r,d-1}^2+k_{r,d}^2)]\Phi_{r,n}(k_r)\Phi_{r,n}(-k_r)\\
S_{ct,int-bf}&=i\frac{A_7\tilde{e}_r\mu^{\epsilon/2}}{\sqrt{N_f}}\sum_{n=1}^3\sum_{j=1}^{N_f}\int dk_r\int dq_r\bar{\Psi}_{r,n,j}(k_r+q_r)\Phi_{r,n}(q_r)\gamma_{d-1}\otimes \sigma_1\Psi_{r,n,j}(k_r)
\end{align}
where we introduced $\Psi_{r,n,j} \equiv \left(\begin{array}{c}\Psi_{r,n,j}^{(-)} \\ \Psi_{r,n,j}^{(+)}\end{array}\right)$ and $\bar{\Psi}_{r,n,j} \equiv \left(\begin{array}{c}\bar{\Psi}_{r,n,j}^{(-)} \\ \bar{\Psi}_{r,n,j}^{(+)}\end{array}\right)^T$.

Based on this effective field theory, we construct Feynman rules
\begin{gather*}
\begin{tikzpicture}[baseline=-0.1cm]
\begin{feynhand}
\vertex (a) at (0,0); \vertex (b) at (1,0); \propagator[fermion] (a) to (b);
\end{feynhand}
\end{tikzpicture}
=\langle \Psi_{r,n,j}(k)\bar{\Psi}_{r,n',j'}(k')\rangle_0=(2\pi)^{d+1}\delta(k-k')\delta_{n,n'}\delta_{j,j'}\left(\begin{array}{cc}-i\frac{\mathbf{\Gamma}\cdot\mathbf{K}_r+\gamma_{d-1}\epsilon_n^{(-)}}{|\mathbf{K}_r|^2+(\epsilon_n^{(-)})^2} & 0 \\ 0 & -i\frac{\mathbf{\Gamma}\cdot\mathbf{K}_r+\gamma_{d-1}\epsilon_n^{(+)}}{|\mathbf{K}_r|^2+(\epsilon_n^{(+)})^2}\end{array}\right)\\
\begin{tikzpicture}[baseline=-0.1cm]
\begin{feynhand}
\vertex (a) at (0,0); \vertex (b) at (1,0); \propagator[boson] (a) to (b);
\end{feynhand}
\end{tikzpicture}
=\langle \Phi_{r,n}(k)\Phi_{r,n'}(k')\rangle_0=(2\pi)^{d+1}\delta(k+k')\delta_{n,n'}\frac{1}{|\mathbf{K}_r|^2+c_r^2(k_{r,d-1}^2+k_{r,d}^2)}\\
\begin{tikzpicture}[baseline=-0.1cm, scale=0.8]
\begin{feynhand}
\vertex (a) at (0,0); \vertex (b) at (1,0); \node at (1,0.3) {$e$}; \vertex (c) at (2,1); \vertex (d) at (2,-1); \propagator[boson] (a) to (b); \propagator[fermion] (b) to (c); \propagator[fermion] (d) to (b);
\end{feynhand}
\end{tikzpicture}
=-i\frac{\tilde{e}_r\mu^{\epsilon/2}}{\sqrt{N_f}}\gamma_{d-1}\otimes \sigma_1,\;\;\;
\begin{tikzpicture}[baseline=-0.1cm, scale=0.8]
\begin{feynhand}
\vertex (a) at (-1,1) {$n$}; \vertex (b) at (1,1) {$n$}; \vertex[ringdot] (c) at (0,0) {}; \vertex (d) at (-1,-1) {$n$}; \vertex (e) at (1,-1) {$n$}; \propagator[boson] (a) to (c); \propagator[boson] (b) to (c); \propagator[boson] (d) to (c); \propagator[boson] (e) to (c);
\end{feynhand}
\end{tikzpicture}
=-\tilde{u}_{1r}\mu^\epsilon,\;\;\;
\begin{tikzpicture}[baseline=-0.1cm, scale=0.8]
\begin{feynhand}
\vertex (a) at (-1,1) {$n$}; \vertex (b) at (1,1) {$n$}; \vertex[dot] (c) at (0,0) {}; \vertex (d) at (-1,-1) {$\bar{n}$}; \vertex (e) at (1,-1) {$\bar{n}$}; \propagator[boson] (a) to (c); \propagator[boson] (b) to (c); \propagator[boson] (d) to (c); \propagator[boson] (e) to (c);
\end{feynhand}
\end{tikzpicture}
=-\tilde{u}_{2r}\mu^\epsilon,\;\;\;
\end{gather*}
for fermion and boson propagators and their interaction vertices and

\begin{gather*}
\begin{tikzpicture}[baseline=-0.1cm]
\begin{feynhand}
\vertex (a) at (0,0); \vertex[crossdot] (b) at (1,0) {}; \vertex (c) at (2,0); \propagator[fermion] (a) to (b); \propagator[fermion] (b) to (c);
\end{feynhand}
\end{tikzpicture}
=-\left(\begin{array}{cc}iA_0k_{r,0}+iA_1\mathbf{\Gamma}_\perp\cdot\mathbf{K}_{r,\perp}+iA_2\gamma_{d-1}\epsilon_n^{(-)}(\frac{A_3}{A_2}v_r) & 0 \\ 0 & iA_0k_{r,0}+iA_1\mathbf{\Gamma}_\perp\cdot\mathbf{K}_{r,\perp}+iA_2\gamma_{d-1}\epsilon_n^{(+)}(\frac{A_3}{A_2}v_r)\end{array}\right)\\
\begin{tikzpicture}[baseline=-0.1cm]
\begin{feynhand}
\vertex (a) at (0,0); \vertex[crossdot] (b) at (1,0) {}; \vertex (c) at (2,0); \propagator[boson] (a) to (b); \propagator[boson] (b) to (c);
\end{feynhand}
\end{tikzpicture}
=-[A_4|q_{r,0}|^2+A_5|\mathbf{Q}_{r,\perp}|^2+A_6c_r^2(q_{r,d-1}^2+q_{r,d}^2)]\\
\begin{tikzpicture}[baseline=-0.1cm, scale=0.8]
\begin{feynhand}
\vertex (a) at (0,0); \vertex[crossdot] (b) at (1,0) {}; \node at (1,0.5) {$e$}; \vertex (c) at (2,1); \vertex (d) at (2,-1); \propagator[boson] (a) to (b); \propagator[fermion] (b) to (c); \propagator[fermion] (d) to (b);
\end{feynhand}
\end{tikzpicture}
=-i\frac{A_7\tilde{e}_r\mu^{\epsilon/2}}{\sqrt{N_f}}\gamma_{d-1}\otimes \sigma_1,\;\;
\begin{tikzpicture}[baseline=-0.1cm, scale=0.8]
\begin{feynhand}
\vertex (a) at (-1,1) {$n$}; \vertex (b) at (1,1) {$n$}; \vertex[crossdot] (c) at (0,0) {}; \vertex (d) at (-1,-1) {$n$}; \vertex (e) at (1,-1) {$n$}; \propagator[boson] (a) to (c); \propagator[boson] (b) to (c); \propagator[boson] (d) to (c); \propagator[boson] (e) to (c);
\end{feynhand}
\end{tikzpicture}
=-A_8\tilde{u}_{1r}\mu^\epsilon,\;\;
\begin{tikzpicture}[baseline=-0.1cm, scale=0.8]
\begin{feynhand}
\vertex (a) at (-1,1) {$n$}; \vertex (b) at (1,1) {$n$}; \vertex[crossdot] (c) at (0,0) {}; \vertex (d) at (-1,-1) {$\bar{n}$}; \vertex (e) at (1,-1) {$\bar{n}$}; \propagator[boson] (a) to (c); \propagator[boson] (b) to (c); \propagator[boson] (d) to (c); \propagator[boson] (e) to (c);
\end{feynhand}
\end{tikzpicture}
=-A_9\tilde{u}_{2r}\mu^\epsilon \;(n\neq \bar{n})
\end{gather*}
for counterterms, respectively. Resorting to these Feynman rules, one can take into account quantum fluctuations perturbatively, where the co-dimensional regularization scheme is utilized.

\subsection{Renormalization group equations}

Correlation functions in terms of bare $\&$ renormalized fermion and boson fields are defined by

\begin{align}
&\langle \Psi_b(k_{b,1})\cdots \Psi_b(k_{b,n_f})\bar{\Psi}_{b}(k_{b,n_f+1})\cdots \bar{\Psi}_{b}(k_{b,2n_f})\Phi_b(q_{b,1})\cdots \Phi_b(q_{b,n_b})\rangle\nonumber \\
&=G_b^{(2n_f,n_b)}(k_{b,i},q_{b,i};v_b,c_b,e_b,u_{1b},u_{2b})\delta^{(d+1)}\Big(\sum_{i=1}^{n_f}(k_{b,i}-k_{b,i+n_f})+\sum_{j=1}^{n_b}q_{b,j}\Big)\\
&\langle \Psi_r(k_{r,1})\cdots \Psi_r(k_{r,n_f})\bar{\Psi}_r(k_{r,n_f+1})\cdots \bar{\Psi}_r(k_{r,2n_f})\Phi_r(q_{r,1})\cdots \Phi_r(q_{r,n_b})\rangle\nonumber \\
&=G_r^{(2n_f,n_b)}(k_{r,i},q_{r,i};v_r,c_r,e_r,u_{1r},u_{2r})\delta^{(d+1)}\Big(\sum_{i=1}^{n_f}(k_{r,i}-k_{r,i+f})+\sum_{j=1}^{n_b}q_{r,j}\Big) ,
\end{align}
respectively.

In order to make the scaling dimension be apparent, we take into account classical scaling (engineering dimension) explicitly as follows
\begin{gather}
\mathbf{K}_r=\mu\tilde{\mathbf{K}},\; k_{r,d-1}=\mu\tilde{k}_{d-1},\; k_{r,d}=\mu\tilde{k}_d \\
\Psi_r=\mu^{-\frac{d+2}{2}}\tilde{\Psi}_r,\; \Phi_r=\mu^{-\frac{d+3}{2}}\tilde{\Phi}_r ,
\end{gather}
where $\mu$ is an energy scale, introduced before. Then, we obtain the renormalization group equation for correlation functions
\begin{align}
G_b^{(2n_f,n_b)}(k_{b,i},q_{b,i};v_b,c_b,e_b,u_{1b},u_{2b})& = Z_\tau Z_{\perp}^{d-2}Z_{\Psi}^{n_f}Z_\Phi^{\frac{n_b}{2}}\mu^{-n_f(d+2)-n_b\frac{d+3}{2}+d+1} \nonumber \\
&\times\tilde{G}_r^{(2n_f,n_b)}(\tilde{k}_{r,i},\tilde{q}_{r,i};v_r,c_r,\tilde{e}_r,\tilde{u}_{1r},\tilde{u}_{2r};\mu) ,
\end{align}
where
\begin{align}
G_r^{(2n_f,n_b)}(k_{r,i},q_{r,i};v_r,c_r,e_r,u_{1r},u_{2r}) &= \mu^{-n_f(d+2)-n_b\frac{d+3}{2}+d+1} \nonumber \\ &\times\tilde{G}_r^{(2n_f,n_b)}(\tilde{k}_{r,i},\tilde{q}_{r,i};v_r,c_r,\tilde{e}_r,\tilde{u}_{1r},\tilde{u}_{2r};\mu) .
\end{align}

Resorting to
\begin{gather}
\frac{d G_b^{(2n_f,n_b)}(k_{b,i},q_{b,i};v_b,c_b,e_b,u_{1b},u_{2b})}{d\ln \mu}=0 ,
\end{gather}
we reformulate the integral form of the renormalization group equation for the correlation function into the differential equation in the following way

\begin{align}\label{CZequation}
&\Big[\sum_{i=1}^{n_f}\Big(z_\tau \tilde{k}_{0}\partial_{0}+z_\perp\tilde{\mathbf{K}}_{\perp,i}\cdot\nabla_{i}+\tilde{k}_{d-1}\partial_{d-1}+\tilde{k}_{d}\partial_{\tilde{k}_d}\Big)+\sum_{i=1}^{n_b}\Big(z_\tau \tilde{q}_{0}\partial_{0}+z_\perp\tilde{\mathbf{Q}}_{\perp,i}\cdot\nabla_{i}+\tilde{q}_{d-1}\partial_{\tilde{q}_{d-1}}+\tilde{q}_{d}\partial_{\tilde{q}_d}\Big)\nonumber \\
&-\beta_{v}\partial_v-\beta_c\partial_c-\beta_e\partial_e-\beta_{u_{1}}\partial_{u_{1}}-\beta_{u_{2}}\partial_{u_{2}}+2n_f\Big(\frac{d+2}{2}-\eta_\Psi\Big)+n_b\Big(\frac{d+3}{2}-\eta_\Phi\Big)-(z(d-1)+2)\Big] \tilde{G}_r^{(2n_f,n_b)}=0 ,
\end{align}
referred to as the Callan-Symanzik equation for the correlation function \cite{RG_Equation_Corr}. Here, we used
\begin{gather}
\frac{dk_{b,0}}{d\ln\mu}=0 \rightarrow \frac{d\tilde{k}_{0}}{d\ln\mu}=-\Big(1+\frac{d\ln Z_\tau}{d\ln\mu}\Big)\equiv -z_{\tau}\tilde{k}_{0}\label{betafunctionsfork1}\\
\frac{d\mathbf{K}_{b,\perp}}{d\ln\mu}=0 \rightarrow \frac{d\tilde{\mathbf{K}}_{\perp}}{d\ln \mu}=-\Big(1+\frac{d\ln Z_\perp}{d\ln \mu}\Big)\equiv -z_\perp\tilde{\mathbf{K}}_\perp\\
\frac{dk_{b,d-1}}{d\ln \mu}=0 \rightarrow \frac{d\tilde{k}_{d-1}}{d\ln \mu}=-\tilde{k}_{d-1}\\
\frac{dk_{b,d}}{d\ln \mu}=0 \rightarrow \frac{d\tilde{k}_{d}}{d\ln \mu}=-\tilde{k}_{d}\label{betafunctionsfork2} .
\end{gather}

Anomalous scaling dimensions for fermion and boson fields are given by
\begin{gather}
\eta_\Psi=\frac{1}{2}\frac{\partial \ln Z_\Psi}{\partial\ln \mu}, \; \; \; \eta_\Phi=\frac{1}{2}\frac{\partial \ln Z_\Phi}{\partial \ln \mu},
\end{gather}
respectively. Beta functions are
\begin{gather}
\beta_v \equiv \frac{d v_r}{d\ln \mu} , \; \; \beta_c \equiv \frac{d c_r}{d\ln \mu},\;\; \beta_e \equiv \frac{d \tilde{e}_r}{d\ln \mu} , \; \; \beta_{u_1} \equiv \frac{d \tilde{u}_{1r}}{d\ln \mu} , \; \; \beta_{u_{2}} \equiv \frac{d \tilde{u}_{1r}}{d\ln \mu} .
\end{gather}

We obtain beta functions based on $\frac{d\ln \mathcal{O}_b}{d\ln\mu}=0$. Suppose $\mathcal{O}_b=\mu^{y_0}Z_1^{y_1} Z_2^{y_2}\cdots Z_{N}^{y_N}\mathcal{O}_r$. Then, the corresponding beta function is given by $\beta_{\mathcal{O}}\equiv \frac{d\mathcal{O}_r}{d\ln \mu}=-\Big(y_0+y_1\frac{d\ln Z_1}{d\ln\mu}+y_2\frac{d\ln Z_2}{d\ln \mu}+\cdots+y_N\frac{d\ln Z_N}{d\ln \mu}\Big)\mathcal{O}_r$. Following this renormalization group equation, we obtain
\begin{gather}
-(\ln Z_0)'+2\eta_{\Psi}+(d-2)(z_\perp-1)+2(z_\tau-1)=0 \label{z0}\\
-(\ln Z_1)'+2\eta_\Psi+(d-1)(z_\perp-1)+(z_\tau-1)=0 \label{z1}\\
-(\ln Z_2)'+2\eta_\Psi+(d-2)(z_\perp-1)+(z_\tau-1)=0\\
-(\ln Z_4)'+2\eta_\Phi+(d-2)(z_\perp-1)+3(z_\tau-1)=0 \label{z4}\\
-(\ln Z_5)'+2\eta_\Phi+d(z_\perp-1)+(z_\tau-1)=0 \label{z5}
\end{gather}
and
\begin{gather}
\beta_v=[(\ln Z_2)'-(\ln Z_3)'] v_r\\
\beta_c=\frac{1}{2}[2\eta_\Phi+(d-2)(z_\perp-1)+(z_\tau-1) -(\ln Z_6)']c_r\\
\beta_e=[-\frac{\epsilon}{2}+\eta_\Phi+2\eta_\Psi+2(d-2)(z_\perp-1)+2(z_\tau-1)-(\ln Z_7)']\tilde{e}_r\\
\beta_{u_1}=[-\epsilon+4\eta_\Phi+3(d-2)(z_\perp-1)+3(z_\tau-1)-(\ln Z_8)']\tilde{u}_{1r}\\
\beta_{u_{2}}=[-\epsilon+4\eta_\Phi+3(d-2)(z_\perp-1) +3(z_\tau-1)-(\ln Z_9)']\tilde{u}_{2r} .
\end{gather}
Here, we used the short-hand notation of $(\ln Z_i)' \equiv \frac{d\ln Z_i}{ d\ln \mu}$. Since $(\ref{z0})-(\ref{z1})$ and $(\ref{z4})-(\ref{z5})$ give two redundant equations, there are actually 9 equations with 9 variables; $z_\tau$, $z_\perp$, $\eta_\Psi$, $\eta_{\Phi}$, $\beta_{v}$, $\beta_c$, $\beta_e$, $\beta_{u_1}$, and $\beta_{u_{2}}$.

Solving these coupled equations, we find renormalization group equations for $z_\tau$, $z_\perp$, $\eta_\Psi$, $\eta_{\Phi}$, $\beta_{v}$, $\beta_c$, $\beta_e$, $\beta_{u_1}$, and $\beta_{u_{2}}$ as follows
\begin{gather}
z_\perp=\Big[1+\frac{1}{2}(F_{e,1}^{(1)}-F_{e,2}^{(1)})\tilde{e}_r+(F_{u_{1},1}^{(1)}-F_{u_{1},2}^{(1)})\tilde{u}_{1r}+(F_{u_{2},1}^{(1)}-F_{u_{2},2}^{(1)})\tilde{u}_{2r}\Big]^{-1}\label{BetaFunctionsFirst}\\
z_{\tau}=z_\perp\Big[1+\frac{1}{2}(F_{e,1}^{(1)}-F_{e,0}^{(1)})\tilde{e}_r+(F_{u_{1},1}^{(1)}-F_{u_{1},0}^{(1)})\tilde{u}_{1r}+(F_{u_{2},1}^{(1)}-F_{u_{2},0}^{(1)})\tilde{u}_{2r}\Big]\\
\eta_{\Psi}=-\frac{1}{2}\Big[z_\perp\Big(1+\frac{1}{2}\tilde{e}_rF_{e,0}^{(1)}+\tilde{u}_{1r}F_{u_{1},0}^{(1)}+\tilde{u}_{2r}F_{u_{2},0}^{(1)}\Big)+2z_\tau-3\Big]+\frac{z_\perp-1}{2}\epsilon\\
\eta_\Phi=-\frac{1}{2}\Big[z_\perp\Big(1+\frac{1}{2}\tilde{e}_rF_{e,4}^{(1)}+\tilde{u}_{1r}F_{u_{1},4}^{(1)}+\tilde{u}_{2r}F_{u_{2},4}^{(1)}\Big) +3z_\tau-4\Big]+\frac{z_\perp-1}{2}\epsilon
\end{gather}
and
\begin{align}
\beta_v&=vz_\perp\Big[\frac{1}{2}\tilde{e}(F_{e,3}^{(1)}-F_{e,2}^{(1)})+\tilde{u_{1}}(F_{u_{1},3}^{(1)}-F_{u_{1},2}^{(1)}) +\tilde{u_{2}}(F_{u_{2},3}^{(1)}-F_{u_{2},2}^{(1)})\Big]\\
\beta_c&=\frac{c_r}{2}\Big[2(1-z_\tau)+z_\perp\Big\{\frac{1}{2}\tilde{e}_r(F_{e,6}^{(1)}-F_{e,4}^{(1)}) +\tilde{u}_{1r}(F_{u_{1},6}^{(1)}-F_{u_{1},4}^{(1)})+\tilde{u}_{2r}(F_{u_{2},6}^{(1)}-F_{u_{2},4}^{(1)})\Big\}\Big]\\
\beta_e&=\tilde{e}_r\Big[1+\frac{1}{2}z_\perp-\frac{3}{2}z_\tau+z_\perp\Big\{\frac{1}{2}\tilde{e}_r\Big(F_{e,7}^{(1)}-F_{e,0}^{(1)}-\frac{1}{2}F_{e,4}^{(1)}\Big)+\tilde{u}_{1r}\Big(F_{u_{1},7}^{(1)}-F_{u_{1},0}^{(1)}-\frac{1}{2}F_{u_{1},4}^{(1)}\Big)\nonumber\\
&+\tilde{u}_{2r}\Big(F_{u_{2},7}^{(1)}-F_{u_{2},0}^{(1)}-\frac{1}{2}F_{u_{2},4}^{(1)}\Big)\Big\}\Big]-\frac{\epsilon}{2}z_\perp \tilde{e}_r\\
\beta_{u_{1}}&=\tilde{u}_{1r}\Big[2+z_\perp-3z_\tau+z_\perp\Big\{\frac{1}{2}\tilde{e}_r(F_{e,8}^{(1)}-2F_{e,4}^{(1)})+\tilde{u}_{1r}(F_{u_{1},8}^{(1)}-2F_{u_{1},4}^{(1)})+\tilde{u}_{2r}(F_{u_{2},8}^{(1)}-2F_{u_{2},4}^{(1)})\Big\}\Big]-\epsilon z_\perp \tilde{u}_{1r}\\
\beta_{u_{2}}&=\tilde{u}_{2r}\Big[2+z_\perp-3z_\tau+z_\perp\Big\{\frac{1}{2}\tilde{e}_r(F_{e,9}^{(1)}-2F_{e,4}^{(1)})+\tilde{u}_{1r}(F_{u_{1},9}^{(1)}-2F_{u_{1},4}^{(1)}) +\tilde{u}_{2r}(F_{u_{2},9}^{(1)}-2F_{u_{2},4}^{(1)})\Big\}\Big]-\epsilon z_\perp\tilde{u}_{2r} . \label{BetaFunctionsFinal}
\end{align}
Here, we used the short-hand notation, given by
\begin{gather}
F_{\mathcal{O},i} \equiv \partial_\mathcal{O}\ln Z_i=\partial_\mathcal{O}\ln\Big(1+\sum_{n=1}^\infty \frac{Z_{i}^{(n)}}{\epsilon^n}\Big)=\partial_\mathcal{O}\sum_{m=1}^{\infty}\frac{(-1)^{m+1}}{m}\Big(\sum_{n=1}^{\infty}\frac{Z_i^{(n)}}{\epsilon^n}\Big)^m\equiv \sum_{n=1}^{\infty}\frac{F_{\mathcal{O},i}^{(n)}}{\epsilon^n} .
\end{gather}

\section{Calculation of Feynman diagrams} \label{1-loopCalculatoin}

\subsubsection{One-loop fermion self-energy correction}

The fermion self-energy correction is given by

\begin{align}
\Sigma^{f(1)}_{n,j}(p)&=-\frac{(\tilde{e}_r)^2\mu^\epsilon}{N_f}\int dk\gamma_{d-1}\otimes \sigma_1G^f_{n}(k+p)\gamma_{d-1}\otimes \sigma_1G^b_{n}(k)\nonumber \\
&=i\frac{\tilde{e}_r^2 }{8\pi^2 c N_f}\frac{1}{\epsilon} \left(\begin{array}{cc}-h_1(c,v)\mathbf{\Gamma}\cdot\mathbf{P}+h_2(c,v)\gamma_{d-1}\epsilon_n^{(+)}(p) & 0 \\  & -h_1(c,v)\mathbf{\Gamma}\cdot\mathbf{P}+h_2(c,v)\gamma_{d-1}\epsilon_n^{(-)}(p)\end{array}\right)
\end{align}
in the one-loop level. From now on, we omit the subscript $r$ in both fermion and boson velocities of $v_{r}$ and $c_{r}$, respectively, for simplicity. As a result, we obtain
\begin{gather}
A_0=A_1=-\frac{\tilde{e}_r^2h_1(c,v)}{8\pi^2 cN_f \epsilon}, \nonumber \\ A_2=-\frac{\tilde{e}_r^2h_2(c,v)}{8\pi^2 cN_f \epsilon}, \;\;\; A_3=-A_2 .
\end{gather}
Here, we used the short-hand notation for
\begin{gather}
h_1(c,v)=\int_0^1dx\sqrt{\frac{x}{xc^2+(1-x)(1+v^2)}},\;\; h_2(c,v)=\int_0^1 dx c^2\sqrt{\frac{x}{[xc^2+(1-x)(1+v^2)]^3}} .
\end{gather}

\subsubsection{One-loop boson self-energy correction}

The boson self-energy correction is given by
\begin{align}
\Pi^{(1)}_{n}(q)&=N_f\Big(\frac{\tilde{e}_r\mu^{\epsilon/2}}{N_f^{1/2}}\Big)^2\int dp \; tr\Big(G_{n}^{f}(p)\gamma_{d-1}\otimes \sigma_1G_{n}^f(p+q)\gamma_{d-1}\otimes \sigma_1\Big)=-\frac{\tilde{e}_r^2|\mathbf{Q}|^2}{16\pi v}\frac{1}{\epsilon}+\cdots .
\end{align}
We recall that both self-interaction vertices of $u_{1}$ and $u_{2}$ do not cause any self-energy corrections in the one-loop level \cite{RG_Equation_Corr}. They result from two loops. Then, we obtain
\begin{gather}
A_4=A_5=-\frac{\tilde{e}_r^2}{16\pi v \epsilon}, \;\;\; A_6=0.
\end{gather}

\subsubsection{One-loop boson-fermion vertex correction}

The boson-fermion vertex correction is given by
\begin{align}
e^{(1)}&=i\frac{(\tilde{e}_r)^3\mu^{3\epsilon/2}}{N_{f}^{3/2}}\int dk \gamma_{d-1}\otimes \sigma_1G_n^{f}(k)
\gamma_{d-1}\otimes \sigma_1G_{n}^{f}(k+q)\gamma_{d-1}\otimes \sigma_1G^b_{n}(k-p)\nonumber \\
&=i\frac{\tilde{e}_r^3}{16\pi^2cN_f^{3/2}\epsilon}h_3(c,v)\gamma_{d-1}\otimes \sigma^1
\end{align}
in the one-loop level, where
\begin{gather}
h_3(c,v)=c\int_0^1dx\int_0^{1-x}dy \frac{2g_1(c,v,x,y)g_2(c,v,x,y)-2v^2(x-y)^2+g_2(c,v,x,y)-v^2g_1(c,v,x,y)}{[g_1(c,v,x,y)g_2(c,v,x,y)-v^2(x-y)^2]^{3/2}}\\
g_1(c,v,x,y)=c^2(1-x-y)+x+y,\;\;\; g_2(c,v,x,y)=c^2(1-x-y)+v^2(x+y) .
\end{gather}
As a result, we find
\begin{gather}
A_7=\frac{\tilde{e}_r^2h_3(c,v)}{16\pi^2cN_f}\frac{1}{\epsilon} .
\end{gather}

\subsubsection{One-loop $u_1$ and $u_{2}$ vertex corrections}

One-loop $u_1$ and $u_{2}$ vertex corrections are essentially the same as those of the $\Phi^{4}$ theory \cite{RG_Equation_Corr}. The $u_{1}$ vertex renormalization is given by
\begin{align}
u_{1}^{(2,0)}&=\frac{\tilde{u}_{1r}^2\mu^{2\epsilon}}{2}\int dk \Big[G_{n}^b(k)G_{n}^b(Q+k) +G_{n}^b(k)G_{n}^b(P+k)+G_n^{b}(k)G_{n}^{b}(K+k)\Big]=\frac{3\tilde{u}_{1r}^2}{16\pi^2 c^2}\frac{1}{\epsilon}\\
u_{1}^{(0,2)}&=\frac{\tilde{u}_{2r}^2\mu^{2\epsilon}}{2}\int dk \sum_{\bar{n}(\bar{n}\neq n)}\Big[G_{\bar{n}}^b(k)G_{\bar{n}}^b(Q+k)+G_{\bar{n}}^b(k)G_{\bar{n}}^b(P+k)+G_{\bar{n}}^b(k)G_{\bar{n}}^{b}(K+k)\Big]=\frac{3\tilde{u}_{2r}^2}{8\pi^2c^2}\frac{1}{\epsilon}
\end{align}
in the one-loop level, where the external momenta are assigned to be $P=p_1+p_2$, $Q=p_1+p_3$, and $K=p_2+p_3$. The superscript $(i,j)$ means $\tilde{u}_{1r}^{i} \tilde{u}_{1r}^{j}$ in the perturbative analysis. As a result, we obtain
\begin{gather}
A_8=\frac{3}{8\pi^2c^2\epsilon}\Big(\frac{\tilde{u}_{1r}}{2}+\frac{\tilde{u}_{2r}^2}{\tilde{u}_{1r}}\Big) .
\end{gather}

Similarly, the $u_{2}$ vertex renormalization is given by
\begin{align}
u_{2}^{(1,1)}&=\frac{\tilde{u}_{1r}\tilde{u}_{2r}\mu^{2\epsilon}}{2}\int dk \Big[G_n^{b}(k)G_{n}^{b}(Q+k)+G_{\bar{n}}^{b}(k)G_{\bar{n}}^{b}(Q+k)\Big]=\frac{\tilde{u}_{1r}\tilde{u}_{2r}}{8\pi^2c^2}\frac{1}{\epsilon}\\
u_{2}^{(0,2)}&=\tilde{u}_{2r}^2\mu^{2\epsilon}\int dk [\frac{1}{2}G_{\bar{\bar{n}}}^b(k)G_{\bar{\bar{n}}}^b(k+Q)+G_{n}(k)G_{\bar{n}}(k+P)+G_{n}(k)G_{\bar{n}}(k+K)]=\frac{5\tilde{u}_{2r}^2}{16\pi^2 c^2}\frac{1}{\epsilon}
\end{align}
in the one-loop level, where the external momentum is assigned to be $Q=p_1-p_3$. As a result, we find
\begin{gather}
A_9=\frac{5\tilde{u}_{2r}+2\tilde{u}_{1r}}{16\pi^2c^2}\frac{1}{\epsilon}.
\end{gather}

\begin{figure}[h]
\begin{tikzpicture}[baseline=-0.1cm]
\begin{feynhand}
\vertex (a) at (-1,1); \vertex (b) at (1,1); \vertex (c) at (-1,-1); \vertex (d) at (1,-1);
\vertex (e) at (-1/2,1/2); \vertex (f) at (1/2,1/2); \vertex (g) at (-1/2,-1/2); \vertex (h) at (1/2,-1/2);
\propagator[boson] (a) to (e); \propagator[boson] (b) to (f); \propagator[boson] (c) to (g);\propagator[boson] (d) to (h);
\propagator[fermion] (e) to (f); \propagator[fermion] (f) to (h); \propagator[fermion] (h) to (g); \propagator[fermion] (g) to (e);
\end{feynhand}
\end{tikzpicture}
\caption{Possible self-interaction boson vertex from the Yukawa coupling in the one-loop order. It turns out to vanish.} \label{quarticYukawa}
\end{figure}
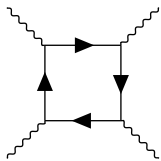

Additionally, there can appear quantum corrections in the one-loop order from the Yukawa vertex as shown in Fig. \ref{quarticYukawa}. However, it turns out to vanish.

Inserting the $A_{n}$ coefficient of the counterterm into the renormalization factor of $Z_{n} = 1 + A_{n}$, we obtain
\begin{gather}
Z_0^{(1)}=Z_1^{(1)}=-\frac{\tilde{e}_r^2 h_1(c,v)}{8\pi^2cN_f}\\
Z_2^{(1)}=-\frac{\tilde{e}_r^2h_2(c,v)}{8\pi^2cN_f}\\
Z_3^{(1)}=-Z_2^{(1)}=\frac{\tilde{e}_r^2h_2(c,v)}{8\pi^2cN_f}\\
Z_4^{(1)}=Z_5^{(1)}=-\frac{\tilde{e}_r^2}{16\pi v}\\
Z_6^{(1)}=0\\
Z_7^{(1)}=\frac{\tilde{e}_r^2h_3(c,v)}{16\pi^2cN_f}\\
Z_8^{(1)}=\frac{3}{8\pi^2c^2}\Big(\frac{\tilde{u}_{1r}}{2}+\frac{\tilde{u}_{2r}^2}{\tilde{u}_{1r}}\Big) \\
Z_9^{(1)}=\frac{5\tilde{u}_{2r}+2\tilde{u}_{1r}}{16\pi^2c^2} .
\end{gather}
where $Z_{n}=\sum_{i=1}^\infty\frac{Z_{n}^{(i)}}{\epsilon^i}$. Introducing these results into the equations of (\ref{BetaFunctionsFirst}) $\sim$ (\ref{BetaFunctionsFinal}), we obtain one-loop beta functions, Eqs. (\ref{1loop_Beta1}) $\sim$ (\ref{1loop_Beta6}).

\end{document}